\documentclass[final,authoryear,5p,times,twocolumn]{elsarticle} 

\usepackage{natbib}
\usepackage{amssymb}
\usepackage{graphicx}
\usepackage{verbatim}
\usepackage[mathscr]{eucal}
\usepackage{epstopdf} 

\usepackage{multirow} 
\usepackage{amsmath}  
\usepackage{float}    
\usepackage{fancyvrb}
\usepackage{cprotect}
\usepackage{lineno}
\newsavebox\verbbox
\journal{Applied Ocean Research}

\begin{document}
\begin{frontmatter}

\author[DTU]{\corref{cor1}Bjarke Eltard Larsen}

\ead{bjelt@mek.dtu.dk}
\cortext[cor1]{Corresponding author}

\author[DTU]{David R. Fuhrman}
\author[DHI]{Johan Roenby}


\title{Performance of interFoam on the simulation of progressive waves}


\address[DTU]{Technical University of Denmark, Department of Mechanical Engineering, Section of Fluid Mechanics, Coastal and Maritime Engineering, DK-2800 Kgs. Lyngby, Denmark}
\address[DHI]{DHI, Department of Port \& Offshore Technology, Agern All\'{e} 5, 2970 H{\o}rsholm}
\begin{abstract}
The performance of \verb|interFoam| (a widely-used solver within the popular open source CFD package \verb|OpenFOAM|) in simulating the propagation of a nonlinear (stream function solution) regular wave is investigated in this work, with the aim of systematically documenting its accuracy. It is demonstrated that over time there is a tendency for surface elevations to increase, wiggles to appear in the free surface, and crest velocities to become (severely) over estimated. It is shown that increasing the temporal and spatial resolution can mitigate these undesirable effects, but that a relatively small Courant number is required.  It is further demonstrated that the choice of discretization schemes and solver settings (often treated as a "black box" by users) can have a major impact on the results. This impact is documented, and it is shown that obtaining a "diffusive balance" is crucial to accurately propagate a surface wave over long distances without requiring exceedingly high temporal and spatial resolutions. Finally, the new code \verb|isoAdvector| is compared to \verb|interFoam|, which is demonstrated to produce comparably accurate results, while maintaining a sharper surface. It is hoped that the systematic documentation of the performance of the \verb|interFoam| solver will enable its more accurate and optimal use, as well as increase awareness of potential shortcomings, by CFD researchers interested in the general CFD simulation of free surface waves.
\end{abstract}


\end{frontmatter}

\section{Introduction}
As a tool to simulate waves \verb|interFoam|, in the widely-used CFD package OpenFOAM (or other solvers build on \verb|interFoam|, e.g.~\verb|waves2Foam| developed by\cite{Jacobsenetal2012}) are becoming increasingly popular. As examples, \verb|interFoam| has been utilized to simulate breaking waves by e.g. \cite{Jacobsenetal2012,Brown2016,Jacobsenetal2014,LupieriContero2015,Higueraetal2013}. It has also been used to simulate wave-structure interaction by e.g. \cite{Higueraetal2013,Chenetal2014,Paulsenetal2014,Huetal2016,Jacobsenetal2015,SchmittElsaesser2015}.

Wave breaking and wave-structure interaction are both very complex phenomena, but \verb|interFoam| has also been utilized to simulate more simple cases, such as the progression of a solitary wave by \cite{Wroniszewsketal2014}, which was suggested as a benchmark to compare to other CFD codes.  The study by \cite{Wroniszewsketal2014} highlighted a problem, that to our knowledge, has gone largely unnoticed in the formal journal literature, namely that the velocity at the crest of the wave is over-predicted relative to the analytical solutions.  This was also highlighted in conference paper \cite{Roenbyetal2017a}, the MSc thesis of \cite{Afshar2010} and the PhD thesis \cite{Tomaselli2016}. A second problem was highlighted in the study by \cite{Paulsenetal2014}, where it was shown that \verb|interFoam| is not capable of maintaining a constant wave height for long propagation distances. They also mentioned, though not going into great detail, that the choice of convection scheme affected this behaviour. The choice of convection scheme was also briefly touched upon by \cite{Wroniszewsketal2014}, who, like \cite{Paulsenetal2014}, utilized an upwind scheme, chosen for stability reasons. A third (again not well described in the literature) problem is the  appearance of wiggles in the air-water interface, as documented by \cite{Afshar2010}. A fourth problem, which has received considerable attention (though not in the context of waves), is the growth of spurious velocities in low density fluid near the interface; see e.g.~\cite{Francoisetal2006,Meieretal2002,Rudman1997,PopinetZaleski1999,Shiranietal2005,Menardetal2007,Tanguyetal2007,GalusinskiVigneaux2008,Hysing2006}. The previous mentioned studies all related the growth of spurious velocities to the surface tension. More recently, however, it should be noted that \cite{Vukcevic2016,Vukcevic2016etal,Wemmenhoveetal2015} demonstrated development of spurious velocities in situations without surface tension.

While a benchmark case as presented in \cite{Wroniszewsketal2014} is, in principal, a good idea many relevant details of the \verb|interFoam| setup were not presented, and this is typically the case in many of the previous mentioned studies.  Such details are quite important, at least from the perspective of benchmarking, as it turns out that the performance of \verb|interFoam| is quite sensitive to the setup (briefly touched upon in \cite{Paulsenetal2014} and \cite{Wroniszewsketal2014} in the choice of convection scheme). Hence, prior to benchmarking \verb|interFoam| or other CFD solvers, it is imperative that an "optimal" (or at least reasonably so) settings be known and utilized.

As the intended audience of the present paper is \verb|OpenFOAM| users, a working knowledge of this software is assumed throughout. To shed light on the general CFD simulation of surface gravity waves, the present study will systematically investigate the performance of \verb|interFoam| on a canonical case involving a simple, intermediately deep, progressive regular wave train. It will demonstrate that taking \verb|interFoam| "out of the box," i.e.~utilizing the standard setup from one of the popular tutorials, will yield quite poor results. After showing the default performance of \verb|interFoam| the sensitivity of \verb|interFoam| to different settings will be investigated. First, a standard sensitive analysis is conducted with respect to the Courant number and mesh resolution. This is done specifically to highlight that commonly-used Courant numbers may not be sufficiently small to accurately simulate gravity waves. Then, utilizing a lower Courant number, different \verb|interFoam| settings will be systematically tested, and finally combined to form a reasonably optimal  set up. More general set up considerations will also be discussed. The recently developed code \verb|isoAdvector| will finally be coupled with \verb|interFoam|, and the performance of \verb|interFoam| (utilizing \verb|isoAdvector| instead of \verb|MULES|) will be compared to the performance of the standard \verb|interFoam| solver.

\section{Model description}

\subsection{Hydrodynamics}

The flow is simulated by solving the continuity equation coupled with momentum equations, respectively given in \eqref{eq:cont} and \eqref{eq:RANS}:
\begin{equation}
\frac{\partial u_i}{\partial x_i} = 0,
\label{eq:cont}
\end{equation}
\begin{equation}
\frac{\partial \rho u_i}{\partial t} + u_j \frac{\partial \rho u_i}{\partial x_j} = -\frac{\partial p^*}{\partial x_i} -g_j x_j \frac{\partial \rho}{\partial x_i} + \frac{\partial}{\partial x_j} \left( 2 \mu S_{ij} \right) +\sigma_T \kappa \frac{\partial \alpha}{\partial x_i},
\label{eq:RANS}
\end{equation}
Here $u_i$ are the mean components of the velocities, $x_i$ are the
Cartesian coordinates, $\rho$ is the fluid density (which takes the constant value $\rho_{\textrm{water}}$ in the water and jumps at the interface to the constant value $\rho_{\textrm{air}}$ in the air phase), $p^*$ is the pressure minus the hydrostatic potential $\rho g_jx_j$, $g_j$ is the gravitational acceleration, $\mu=\rho\nu$ is the dynamic molecular viscosity ($\nu$ being the kinematic viscosity), and $S_{ij}$ is the mean strain rate tensor given by
\begin{equation}
S_{ij} = \frac{1}{2} \left( \frac{\partial u_i}{\partial x_j} + \frac{\partial u_j}{\partial x_i}\right).
\end{equation}

The last term in equation \eqref{eq:RANS} accounts for the effect of surface tension, $\sigma_T$, where $\kappa$ is the local surface curvature and $\alpha$ is the so-called indicator field introduced for convenience, which takes value 0 in air and 1 in water. It can be defined in terms of the density as
\begin{equation}
	\alpha = \frac{\rho-\rho_{\textrm{air}}}{\rho_{\textrm{water}}-\rho_{\textrm{air}}}.
\end{equation}
We assume that any intrinsic fluid property, $\Phi$, can be expressed in terms of $\alpha$ as
\begin{equation}
\Phi = \alpha \Phi_{water} + (1-\alpha)\Phi_{air}.
\end{equation}
The evolution of $\alpha$ is determined by the continuity equation, which in terms of $\alpha$ reads
\begin{equation}
\frac{\partial \alpha}{\partial t} + \frac{\partial \alpha u_j }{\partial x_j} = 0.
\label{eq:alpha1}
\end{equation}
In \verb|interFoam| the numerical challenge of keeping the interface sharp is addressed using a numerical interface compression method and limiting the phase fluxes based on the "Multidimensional universal limiter with explicit solution" (\verb|MULES|) limiter. Numerical interface compression is obtained by adding a purely heuristic term to equation \eqref{eq:alpha1}, such that it attains the form
\begin{equation}
\frac{\partial \alpha}{\partial t} + \frac{\partial \alpha  u_j }{\partial x_j} + \frac{ \partial  }{\partial x_j}\left(\alpha (1-\alpha)u^{r}_j \right) = 0.
\label{eq:alpha1num}
\end{equation}
Here $u_j^r$ is a modelled relative velocity used to compress the interface. For more details on the numerical implementation, the reader is referred to \cite{Deshpandeetal2012}.

All simulations are performed utilizing \verb|OpenFOAM| version \verb|foam-extend 3.2|. The authors are aware of a "new" \verb|MULES| algorithm (not present in the extend versions) in newer versions from \verb|OpenFOAM-2.3.0|, and also of the new commit support for \verb|Crank-Nicolson| on the time integration of $\alpha$. Therefore the base case to be presented later, was also simulated utilizing a newer version of the standard \verb|OpenFOAM|, namely \verb|OpenFOAM-3.0.1|. We were unable to produce significantly different results with these newer versions as compared to our simulations with \verb|foam-extend 3.2|, hence the base performance demonstrated in what follows is likewise expected to be representative of newer versions. 

\subsection{Boundary and initial conditions}
For this study a simple base case of a regular propagating wave will be simulated with various numerical settings. The quality of the simulated wave will be assessed through comparison with the analytical solution in terms of surface elevations and velocity profiles. We use a so-called stream function wave from \cite{RieneckerFenton1981}, initialized with \verb|waves2Foam| developed by \cite{Jacobsenetal2012},  with a period $T=2$ s and wave height $H=0.125$ m at a water depth of $h=0.4$ m. This gives $kh=0.66$ and $H/h=0.31$, which indicates that the simulated wave is non-linear and at intermediate depth, with $k$ being the wave number. The stream function solution can be considered as a numerically exact wave solution based on nonlinear potential flow equations. The properties have been selected to correspond to the incoming wave in the well-known spilling breaker experiment of \cite{TingKirby1994}.
For all simulations the wave will be propagated through a domain which is exactly one wave length long and two water depths high with cyclic periodic boundary conditions on the sides. Unless stated otherwise the domain is discretised into cells having an aspect ratio of 1 with the number of cells per wave height $N=H/\Delta y=12.5$, resulting in cells with $\Delta x=\Delta y=0.01$ m. This results in a two dimensional domain with 379$\times$80 cells. At the bed a slip condition is utilized in accordance with potential flow theory. At the top the \verb|pressureInletOutletVelocity| is used. This means that there is a zero gradient condition except on the tangential component which has a value of zero.

\section{interFoam settings}
In this section the default numerical settings for our simulations, as well as a general description of \verb|OpenFOAM|'s discretization practices, are presented. Our base numerical settings will be those found in the popular \verb|damBreak| tutorial shipped with \verb|foam-extend-3.2|. With this starting point we will change various settings to investigate their effect on the quality of the numerical solution. More specifically, we copy the \verb|controlDict|, \verb|fvSchemes| and \verb|fvSolution| files directly from the \verb|damBreak| tutorial. In the \verb|constant| directory the mesh and the physical parameters of the case are specified: $\rho_{\textrm{water}}$=1000 kg/m$^3$, $\rho_{\textrm{air}}=$ 1.2 kg/m$^3$, $\nu_{\textrm{water}}=1 \cdot 10^{-6}$ m$^2$/s, $\nu_{\textrm{air}}= 1.45 \cdot 10^{-5}$ m$^2$/s, and $\sigma_T = 0.0 $ N/m (i.e.~no surface tension).
We note that the analytic stream function solution does not take into account the presence of air, nor the effect of viscosity or surface tension. With the chosen wave parameters and boundary conditions (e.g.~no slip at the bed) the physics are dominated by inertia and gravity. With a density rate of $\rho_{\textrm{water}}/\rho_{\textrm{air}} \sim 833$, the air will behave like a ``slave fluid'' moving passively out of the way for the water close to the surface. To confirm the insignificance of the physical viscosity in our setup, we have compared simulations with these set to their physical values and to zero, and confirmed that this had no effect on our results. We have also performed simulations with $\rho_{\textrm{air}}=0.1$ kg/m$^3$ and $\rho_{\textrm{air}}=10$ kg/m$^3$. This had almost no effect in the short term, but had some effect for long propagation distances. Increasing the density made the air behave less like a ``slave fluid'' and slowed the propagation of the wave. Decreasing the density created larger air velocities, but did not alter the wave kinematics significantly. We have confirmed that switching the surface tension between zero and its physical value ($\sigma_T=0.07$ N/m) had next to no effect on our simulation results, as expected in the gravity wave regime. Finally, the simulations are performed without turbulence, as the results are intended to be compared with the idealized stream function (potential flow) solution. 

The  \verb|OpenFOAM| case setup is contained in a file called \verb|controlDict| which, among others things, controls the time stepping method. The schemes used to discretize the different terms in the governing equations are specified in the \verb|fvSchemes| file, and the file \verb|fvSolution| contains various settings for the linear solvers and for the solution algorithm. In Table \ref{tab:damSetup} the essential parameters for the base set up from these three files are indicated. 
\begin{table*}[t]
\centering
\caption{Base setup from the damBreak tutorial}

\begin{tabular}{ccl}\hline

\verb|controlDict|  &  Scheme/Value \\ 
\hline
\verb|adjustTimeStep| & \verb|true|\\

\verb|maxCo|&0.5\\

\verb|maxAlphaCo|&0.5\\
\hline
\verb|fvSchemes|\\
\hline
\verb|ddt|& \verb|Euler|\\
\verb|grad|& \verb|Gauss Linear|\\
\verb|div(rho*phi,U)|& \verb|Gauss LimitedLinearV 1|\\
\verb|div(phi,alpha1)|& \verb|Gauss VanLeer01|\\
\verb|div(phirb,alpha1)|& \verb|Gauss interfaceCompression|\\
\verb|laplacian|& \verb|Gauss linear corrected|\\
\verb|interpolation|& \verb|linear|\\
\verb|snGrad|& \verb|corrected|\\
\hline
\verb|fvSolution|\\
\hline
\verb|pcorr(solver,prec,tol,relTol)|&\verb|PCG, DIC|, 1e-10, 0\\
\verb|pd(solver,prec,tol,relTol)|&\verb|PCG, DIC|, 1e-07, 0.05\\
\verb|pdFinal(solver,prec,tol,relTol)|&\verb|PCG, DIC|, 1e-07, 0\\
\verb|U(solver,prec,tol,relTol)|&\verb|PBiCG, DILU|, 1e-06, 0\\
\verb|cAlpha|&1\\
\verb|momentumPredictor|&\verb|yes|\\
\verb|nOuterCorrectors|&1\\
\verb|nCorrectors|&4\\
\verb|nNonOrthogonalCorrectors|&0\\
\verb|nAlphaCorr|&1\\
\verb|nAlphaSubCycles|&2

\end{tabular}

\label{tab:damSetup}
\end{table*}
The most important details of the scheme and solver choices presented in Table \ref{tab:damSetup} will be described in the following. For descriptions of the remaining settings, the reader is referred to the \verb|OpenFOAM| user guide and programmers guides in \cite{Greenshields2015,Greenshields2016}.

\subsection{controlDict}
In this subsection the most important \verb|controlDict| settings are presented. The time step can be specified either as \verb|fixed|, such that the user defines the size of the time step, or as \verb|adjustable|. In the latter case the time step is adjusted such that a maximum Courant number $Co=u_i \Delta t/ \Delta x_i$ or a maximum \verb|AlphaCo| (The Courant number in interface cells) is maintained at all times. 
In the \verb|damBreak| tutorial an adjustable time step is used with $Co=0.5$, hence this value will be utilized initially.

\subsection{fvSchemes}
In this subsection some of the discretisation schemes are presented to aid in the understanding of the forthcoming analysis.
The \verb|ddt| scheme specifies how the  time derivative $\partial/\partial t$ is handled in the momentum equations. Available in \verb|OpenFOAM| are: \verb|steadyState|, \verb|Euler|, \verb|Backwards| and \verb|CrankNicolson|. In this study, \verb|steadyState| is naturally disregarded as the simulations are unsteady. The \verb|Euler| scheme corresponds to the first-order forward Euler scheme, whereas \verb|Backward| corresponds to a second-order, \verb|OpenFOAM| implemented time discretization scheme, which utilizes the current and two previous time steps. The \verb|CrankNicolson| (CN) scheme includes a blending factor $\psi$, where $\psi=1$ corresponds to pure (second-order accurate) CN and $\psi=0$ corresponds to pure \verb|Euler|. This blending factor is introduced to give increased stability and robustness to the CN scheme.

In the finite volume approach used in \verb|OpenFOAM|, the convective terms in the mass \eqref{eq:alpha1num} and momentum \eqref{eq:RANS} equations  are integrated over a control volume, and afterwards the Gauss theorem is applied to convert the integral into a surface integral:
\begin{equation}
\int_V \nabla \cdot \left(\phi u\right) dV = \oint_S \phi \left(n \cdot u\right) dS \approx \sum_f \phi_f F_f,
\end{equation}
where $\phi(x,t)$ is the field variable, $\phi_f$ is an approximation of the face averaged field value and $F_f=s_f \cdot u_f$ is the face flux, with $s_f$ being the face area vector normal to the face pointing out of the cell. $\phi_f$ can be determined by interpolation, e.g. using central or upwind differencing. Central differencing schemes are second order accurate, but can cause oscillations in the solution. Upwind differencing schemes are first order accurate, cause no oscillations, but can be very diffusive. 

\verb|OpenFOAM| includes a variety of total variation diminishing (TVD) and normalized variable diagram (NVD) schemes aimed at achieving good accuracy while maintaining boundedness. TVD schemes calculate the face value $\phi_f$ by utilizing combined upwind and central differencing schemes according to
\begin{equation} 
\phi_f  = (1-\Gamma) \phi_{f,UD}+\Gamma \phi_{f,CD0}
\end{equation} 
where $\phi_{f,UD}$ is the upwind estimate of $\phi_f$, $\phi_{f,CD}$ is the central differencing estimate of $\phi_f$. $\Gamma$ is a blending factor, which is a function of the variable $r$ representing the ratio of successive gradients,
\begin{equation}
r=2 \frac{d \cdot \left(\nabla \phi\right)_P}{\phi_N-\phi_P}.
\end{equation}
Here $d$ is the vector connecting the cell centre $P$ and the neighbour cell centre $N$. In NVD-type schemes the limiter is formulated in a slightly different way.
In the \verb|damBreak| tutorial base setup the TVD scheme is utilized by specifying the keyword \verb|limitedLinearV 1| for the momentum flux, \verb|div(rho*phi,U)|, and \verb|vanLeer01| for the mass flux, \verb|div(phi,alpha1)|, where the keyword \verb|phi| means face flux. With the \verb|limitedLinear|  scheme $\Gamma=max\left[min\left(2r/k,1\right),0\right]$,  where $k$ is an input given by the user, in this case $k=1$. When using the scheme for vector fields a "V" can be added to the TVD schemes, which changes the calculation of $r$ to take into account the direction of the steepest gradients. The \verb|vanLeer| scheme calculates the blending factor as $\Gamma=(r+|r|)/(1+|r|)$. The \verb|01| added after the TVD scheme name means that $\Gamma$ is set to zero if it goes out of the bounds 0 and 1, thus going to a pure upwind scheme to stabilize the solution. The other available TVD/NVD schemes differ in their definition of $\Gamma$ and resulting degree of diffusivity. Since $r$ depends on the numerically calculated gradient of $\phi$, the choice of gradient scheme can also play an important role. In general the gradients are calculated utilizing a Gauss linear scheme, but this might lead to unbounded face values, and therefore gradient limiting can be applied. As an example the gradient scheme can be specified as \verb|Gauss faceMDLimited|. The keyword \verb|face| or \verb|cell| specifies whether the gradient should be limited base on cell values or face values and the keyword \verb|MD| specifies that it should be the gradient normal to the faces. In addition to the linear choice of gradient schemes there also exists a least square scheme as well as a fourth order scheme.

The \verb|laplacian| scheme specifies how the Laplacian in the pressure correction equation within the \verb|PISO| algorithm, as well the third term on the right hand side of equation \eqref{eq:RANS}, should be discretized. It requires both an interpolation scheme for the dynamic viscosity, $\mu$, and a surface normal gradient scheme \verb|snGrad| for $\nabla u$. Often a \verb|linear| scheme is used for the interpolation of $\mu$ and the proper choice of surface normal gradient scheme depends on the orthogonality of the mesh. Besides being used in the Laplacian, the \verb|snGrad| is also used to evaluate the second and fourth term on the right hand side of equation \eqref{eq:RANS}. Often a linear scheme will be used, with or without orthogonality correction. Another option is to use a fourth order surface normal gradient approximation. Finally, the interpolation scheme determines how values are interpolated from cell centres to face centres. 

\subsection{fvSolution}
In the \verb|fvSolutions| file the iterative solvers, solution tolerances and algorithm settings are specified. The available iterative solvers are preconditioned (bi-)conjugate gradient solvers denoted \verb|PCG/PBiCG|, a \verb|smoothSolver|, generalised geometric-algebraic multi-grid, denoted \verb|GAMG|, and a \verb|diagonal| solver. Each solver can be applied with different preconditioners and the smooth solver also has several smoothing options. The \verb|GAMG| solver works by generating a quick solution on a coarse mesh consisting of  agglomerated cells, and then mapping this solution as the initial guess on finer meshes to finally obtain an accurate solution on the simulation mesh. The different preconditioners and smoothers will not be discussed here, but \cite{Greenshields2015, Greenshields2016} can be consulted for additional details.

In addition to the solver choices the \verb|PISO|, \verb|PIMPLE| and \verb|SIMPLE| controls are also given in the \verb|fvSolution| file.  The \verb|cAlpha| keyword controls the magnitude of the numerical interface compression term in equation \eqref{eq:alpha1num}. \verb|cAlpha|  is usually set to 1 corresponding to a ``compression velocity'' of the same size as the flow velocity at the interface. The \verb|momentumPredictor| is a switch specifying enabling activation/deactivation of the predictor step in the \verb|PISO| algorithm. The parameter, \verb|nOuterCorrectors| is the number of outer correctors used by the \verb|PIMPLE| algorithm and specifies how many times the entire system of equations should be solved within one time step. To run the solver in ``PISO mode'' we set \verb|nOuterCorrectors| to 1. The parameter \verb|nCorrectors| is the number of pressure corrector iterations in the \verb|PISO| loop. The parameter \verb|nAlphaSubCycles| enables splitting of the time step into \verb|nAlphaSubCycles| in the solution of the $\alpha$ equation \eqref{eq:alpha1num}. Finally, the parameter \verb|nAlphaCorr|, specifies how many times the \verb|alpha| field should be solved within a time step, meaning that first the alpha field is solved for, and this new solution is then used in solving for the alpha field again. 
\section{Results and discussion}
In this section the simulated results involving the propagation of the regular stream function wave will be presented and discussed for various settings.
\subsection{Perfomance of interFoam utilizing the damBreak settings}
First, the "default" performance of \verb|interFoam| in the progression of the stream function wave is presented, utilizing the settings from the \verb|damBreak| tutorial. The setup utilized here will be considered as the base setup, and the remainder of the simulations in this study will utilize this base setup with minor adjustments. 

\begin{figure}[ht]
	\centering
    \includegraphics[trim=0cm 0cm 0cm 0.5cm, clip=true, scale=1]{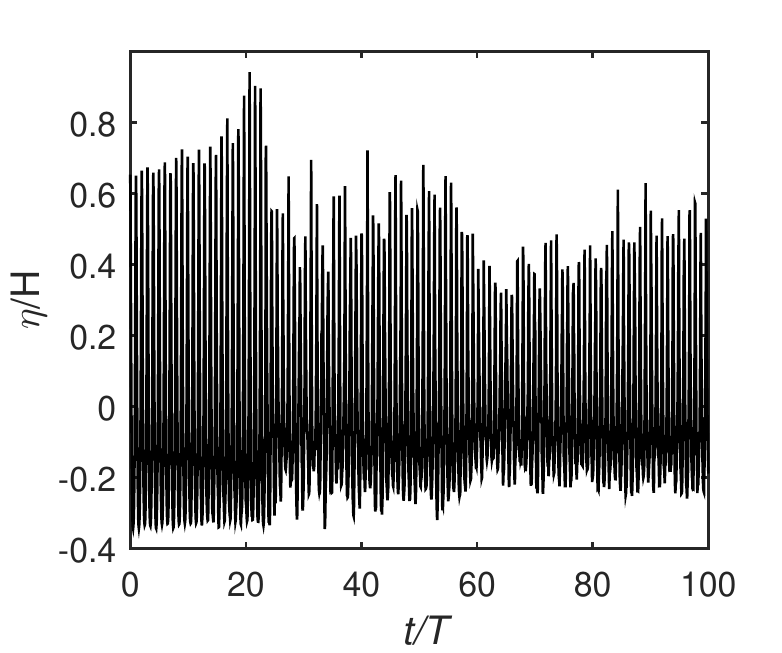}
    \caption{Surface elevation for the propagating wave utilizing the damBreak setup}
    \label{fig:damElevation}
 \end{figure}

\begin{figure}[ht]
	\centering
    \includegraphics[scale=0.65]{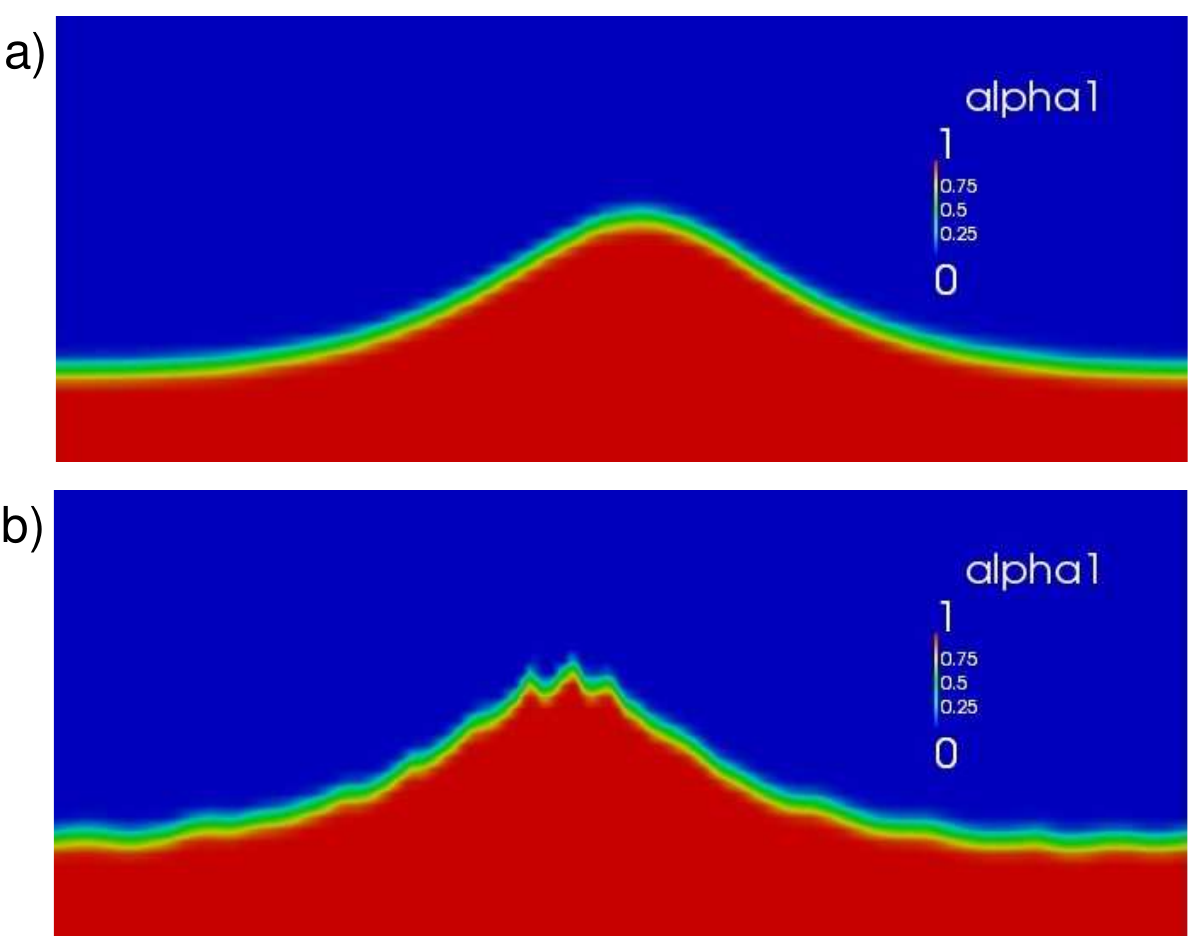}
    \caption{Snapshot at a) $t=5.5 T$ and b) $t=16.25 T$,  illustrating the appearance of small wiggles in the crest after sufficiently long propagation}
    \label{fig:SurfacedamSetup}
 \end{figure}

Starting from the analytical stream function solution imposed as an initial condition (utilizing the \verb|waves2Foam| toolbox of \cite{Jacobsenetal2012}), the simulation is performed for 200 s (corresponding to 100 periods). This is sufficiently long to  highlight certain strengths and  problems of \verb|interFoam|. Results are sampled at the cyclic boundary 20 times per period. In Figure \ref{fig:damElevation} the surface elevation time series is shown. Quite noticeably, even though the depth is constant, the wave height immediately starts to increase, and this continues until the wave at some point (approximately at $t=20 T$) breaks. This rather surprising result demonstrates the potentially poor performance of \verb|interFoam|, as the wave does not come close to maintaining a constant form. A similar result has been shown in \cite{Afshar2010}.
 A feature that seems to contribute, though is not solely responsible for, the un-physical steepening of the wave, is small "wiggles" on the interface. These are illustrated in Figure \ref{fig:SurfacedamSetup} where a snapshot of the wave is seen after approximately five and 16 periods. The vertical axes are exaggerated to highlight the presence of the wiggles. As the wave propagates these wiggles emerge, continue to grow and sometimes merge, hence contributing to the steepening of the wave, which ultimately breaks. The cause of the wiggle feature will be discussed in Section \ref{sec:Schemes}.

\begin{figure*}[ht]
	\centering
    \includegraphics[trim=0cm 0cm 0cm 0.5cm, clip=true, scale=1]{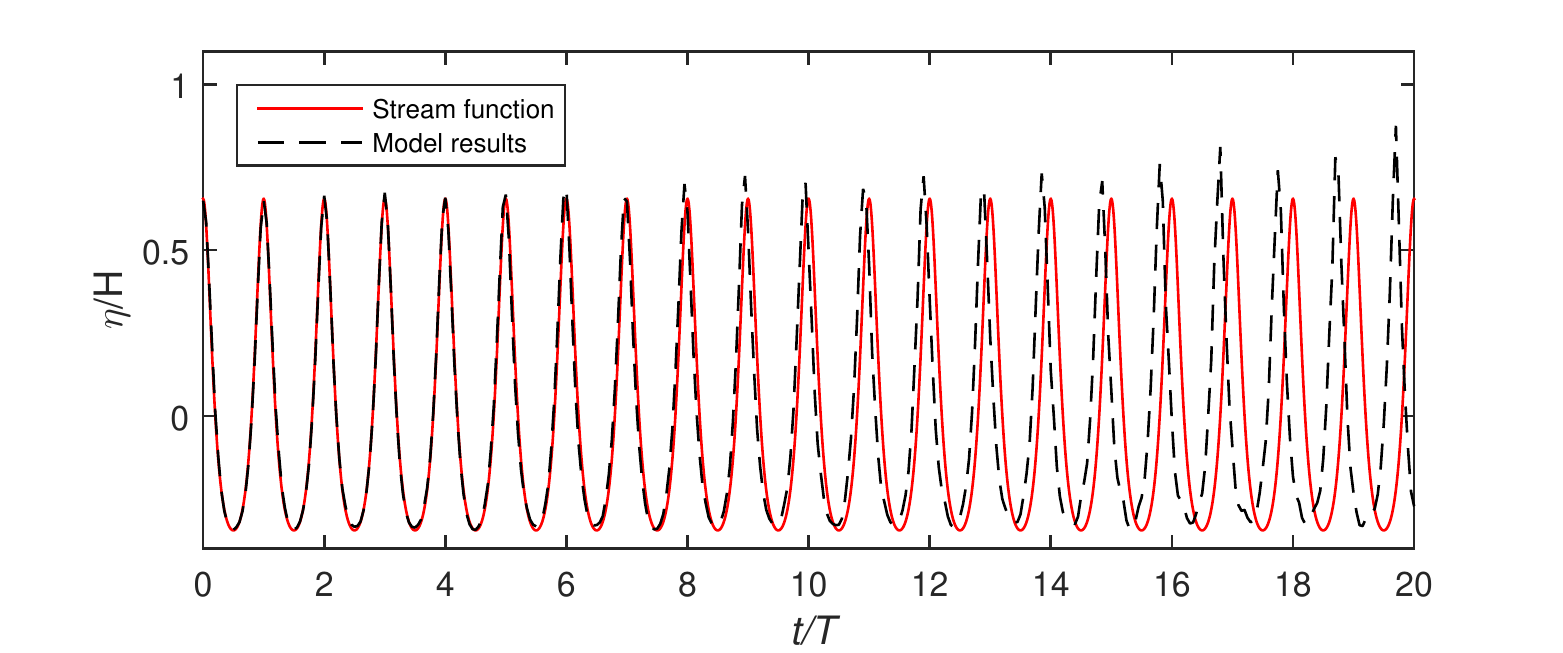}
    \caption{Surface elevation for the propagating wave utilizing the damBreak setup}
    \label{fig:damElevationCompare}
 \end{figure*}

While propagating, in addition to steepening, the celerity is also increasing compared to the analytical stream function solution, resulting in a phase error. To demonstrate this the surface elevation for the first 20 periods is compared with the stream function solution in Figure \ref{fig:damElevationCompare}. Here it is quite evident that significant phase errors occur after approximately propagating for 10 periods, where the simulated results start to lead the analytical solution. This corresponds approximately to the time where over-steepening is apparent, hence the phase error may be attributed to the un-physical increase in the nonlinearity of the wave.

 
\begin{figure}[ht]
	\centering
    \includegraphics[scale=1]{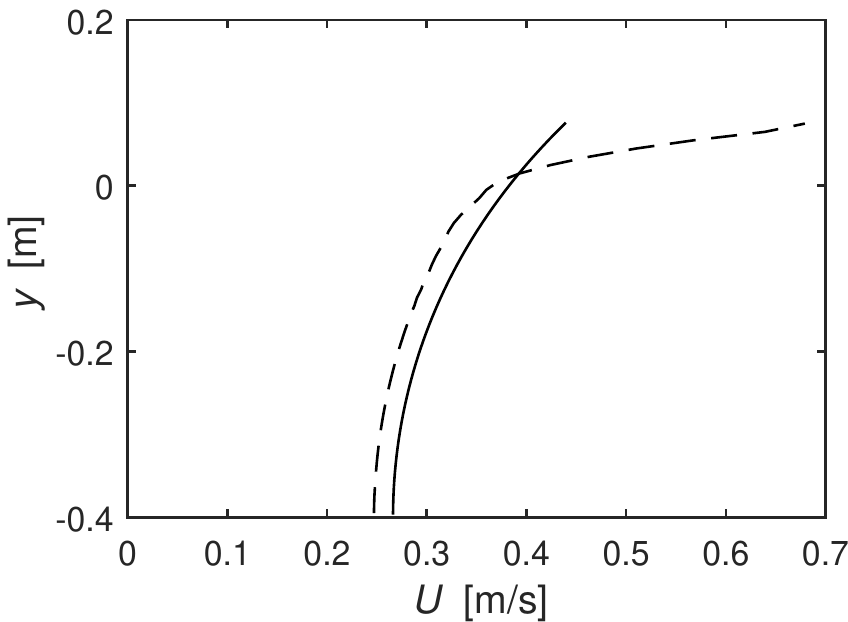}
    \caption{Modelled velocity distribution beneath the crest (- -) and stream function solution, (-) at $t=5 T$.}
    \label{fig:velocitydamSetup}
 \end{figure}

Also of great interest is the velocity profile beneath the propagating wave, as velocity kinematics often form the basis for force calculations on coastal or offshore structures, while also influencing e.g. bed shear stresses and hence sediment transport predictions (in simulations where the boundary layer is also resolved). In Figure \ref{fig:velocitydamSetup} the velocity profile directly beneath the crest of the wave after five periods is shown together with the analytical stream function solution. It should be noted that the velocity here, and in future results, is taken as $U=u_1 \alpha$, and it is only shown from the bed until the height where it reaches its maximum value. This is done to capture the velocity all the way to the crest of the wave and not merely to a predefined height (as just shown, the wave height increases). Furthermore, this formulation also includes the velocities at cells containing a mixture of air and water, which is desirable, as some diffusion of the interface is seen. 

As seen in Figure \ref{fig:velocitydamSetup}, the velocity beneath the crest is underestimated close to the bed and, especially near the free surface, is severely overestimated. This is despite the fact that the wave has still reasonably maintained its shape up to this time, see Figure \ref{fig:SurfacedamSetup}a and \ref{fig:damElevationCompare}. 
This over-predicted crest velocity, in addition to the steepening of the wave, also likely contributes to the wave breaking. The overestimation of crest velocities in regular waves  by \verb|interFoam| has, to our knowledge, gone almost un-recognized in the journal literature. It is recorded in \cite{Wroniszewsketal2014} in the propagation of a solitary wave and in \cite{Roenbyetal2017a} as well as in the MSc thesis of \cite{Afshar2010} and the PhD thesis of \cite{Tomaselli2016}. 
The overesitmation of the crest velocity is believed to arise from an imbalance in the discretized momentum equation near the interface. As the wave propagates the increase in crest velocity becomes continually worse, and in addition to the imbalance in the momentum equation near the free surface, the steepening of the wave also contributes to this increase.  
 
Finally, though not shown herein for brevity, we note that regions of high air velocities were seen to develop just above the free surface and in the mixture cells. Such spurious velocities have elsewhere been attributed to surface tension effects,  see e.g. \cite{Deshpandeetal2012}, but the spurious velocities found in these simulation are clearly of a different nature as the surface tension is turned off. The main challenge leading to this behavior is that when the water/air density ratio is high, even small erroneous transfers of momentum across the interface from the heavy to the light fluid will cause a large acceleration of the light fluid, as also discussed by \cite{Vukcevic2016,Vukcevic2016etal,Wemmenhoveetal2015}. The resulting large air velocities may then be subsequently diffused back across the interface into the water, the degree to which will be discussed in Section \ref{sec:Schemes}.

\subsection{Effect of the Courant number, $Co$}
With the poor performance previously shown using the default \verb|damBreak| settings, two natural places to attempt improvement in the solution would be in the temporal and spatial resolutions. In this section the effect of the temporal resolution will be investigated by varying $Co$.

\begin{figure*}[ht]
	\centering
    \includegraphics[scale=1]{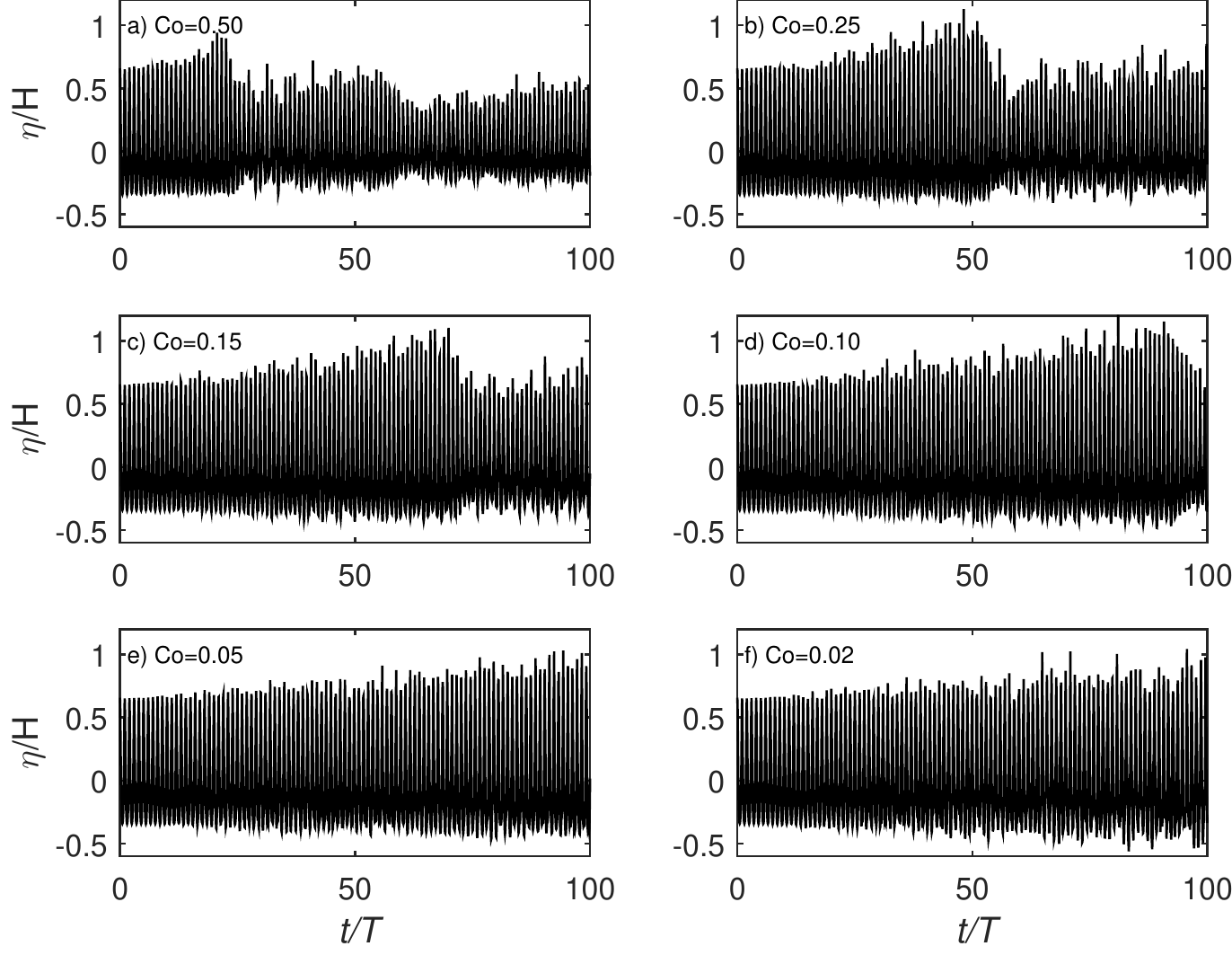}

    \cprotect\caption{Simulated surface elevation as a function of time for six different Courant numbers (Main fixed parameters: $N=12.5$, \verb|ddt-Euler, grad-Gauss Linear, div(rho*phi,U)- Gauss LimitedLinearV 1, laplacian-Gauss linear| \verb| corrected|, $c_\alpha =1$).}
    \label{fig:CoElevation}
 \end{figure*}

Figure \ref{fig:CoElevation} shows the surface elevation as a function of time for six different values of $Co$. From this it is evident that lowering $Co$ has a significant impact on \verb|interFoam|'s performance. However, even with $Co=0.02$ \verb|interFoam| it is not capable of keeping the wave shape for 100 periods as the wave heights are still seen to increase. Up until 20 wave periods the wave height is close to constant when using $Co \leq 0.15$. 
The wave is still leading the analytical stream function solution and in general lowering $Co$  reduces the overestimation of the wave celerity as can be seen in table \ref{tab:PhaseShift} where the phase-shift at $t=25T$ is shown for the six different values of $Co$.
The phase shift is calculated as $\phi_{shift}=(t_{peak}-t_{analytical})/T \cdot 360^{\circ}$, where $t_{peak}$ is the time where the crest of the wave passes the sampling position, and $t_{analytical}$ is the time where the stream function solution should have passed the sampling position.  
 \begin{table}[ht]
\centering
\caption{Phase-shift at $t=25 T$.}
\begin{tabular}{cccccccl}\hline
$Co$ &0.02&0.05&0.10&0.15&0.25&0.50 \\
$\phi_{shift}$ [$^{\circ}$]&0.0&0.0&-18&-36&-72&-198
\end{tabular}
\label{tab:PhaseShift}
\end{table}

\begin{figure}[ht]
	\centering
    \includegraphics[scale=1]{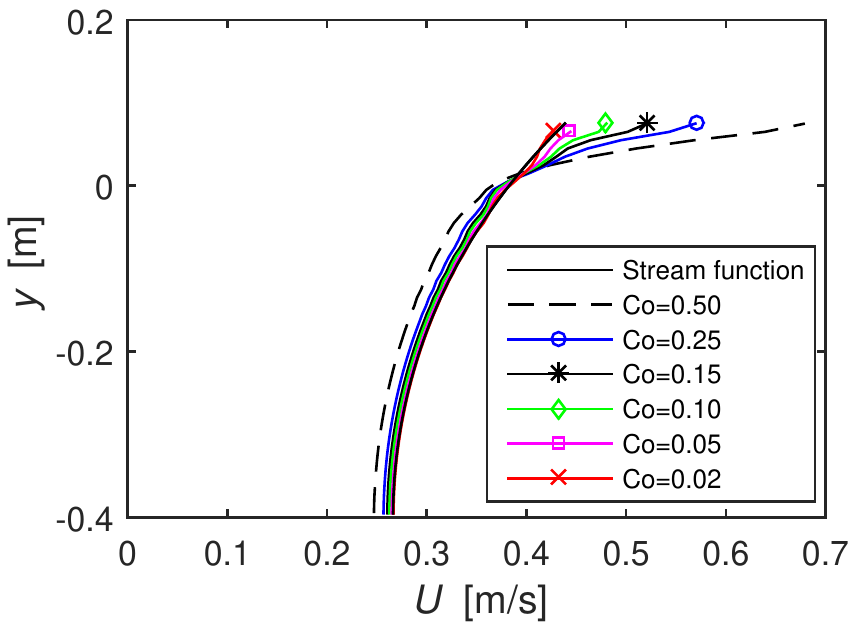}

    \cprotect\caption{Velocity distribution beneath the crest at $t=5 T$ for various Courant numbers (Main fixed parameters: $N=12.5$, \verb|ddt-Euler, grad-Gauss Linear, div(rho*phi,U)-| \verb|Gauss LimitedLinearV 1, laplacian-Gauss linear corrected| $c_\alpha =1$). }
    \label{fig:velocityCo}
 \end{figure}

Figure \ref{fig:velocityCo} shows the velocity profiles beneath the crest at $t=5T$ for the six different values of $Co$ together with the stream function solution, similar to Figure \ref{fig:velocitydamSetup}. It can be seen that as $Co$ is lowered the simulated velocity profiles become closer to the analytical solution. The reason for this is probably two-fold. First, lowering $Co$ delays the presence and growth of the interface wiggles and thus also the steepening of the wave. Second, any inconsistent treatment of the force balance near the free surface is substantially limited by the small time step as it reduces e.g. the error committed in linearising the convective term $u_j (\partial \rho u_i/\partial x_j)$. The importance of keeping a low time step in \verb|interFoam| when doing two-phase simulations has also been highlighted by \cite{Deshpandeetal2012} in the context of surface tension dominated flows, where it was shown that a small time step is crucial for limiting the growth of spurious velocities. Even though the present inertia dominated situation is different from the analysis of \cite{Deshpandeetal2012}, the solution to minimize the interface imbalance by limiting the time step still seems to hold.  
 
\begin{figure}[ht]
	\centering
    \includegraphics[scale=1]{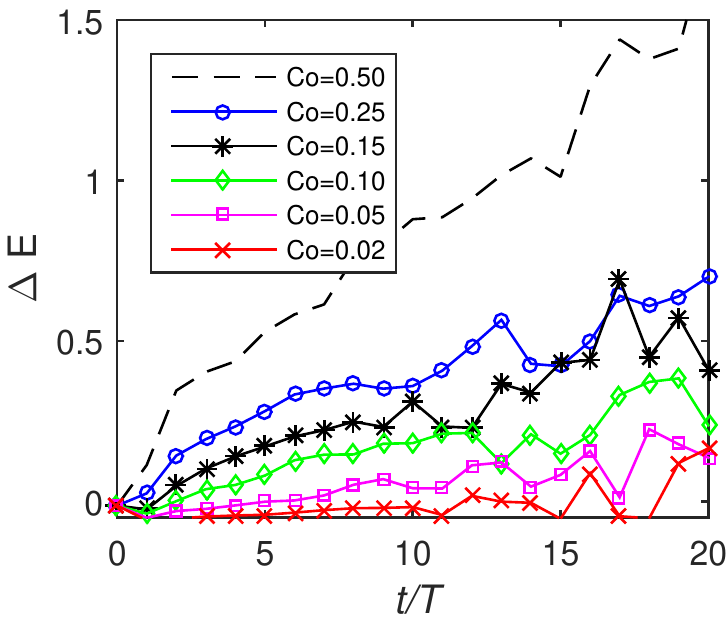}
   \cprotect\caption{Error in the maximum crest velocity as a function of periods (Main fixed parameters: $N=12.5$, \verb|ddt-Euler, grad-Gauss Linear, div(rho*phi,U)-|\verb|Gauss| \verb|LimitedLinearV 1, laplacian-Gauss linear corrected| $c_\alpha =1$)..}
    \label{fig:ErrorCo}
 \end{figure}

In addition to the velocity profiles depicted in Figure \ref{fig:velocityCo}, it is also of interest to see how the overestimation of the crest velocity evolves in time. Therefore, in Figure \ref{fig:ErrorCo} the error in the crest velocity calculated as 
\begin{equation}
\Delta E =\frac{max(U)-U_{analytical}}{U_{analytical}}
\end{equation}
is shown for each of the six values of $Co$ considered. 
Regardless of $Co$, the overestimation of the crest velocity is apparent and grows in time. From Figure \ref{fig:ErrorCo} it can be seen that even with a relatively small $Co$, e.g.  $Co=0.15$, after only propagating five periods, the crest velocity is approximately 17$\%$ larger than the analytical. It thus seems that, what is generally viewed as a rather "low" $Co$, is still not sufficiently small to accurately simulate surface waves. In contrast, the error in the crest velocity for the case with $Co=0.05$ is only $0.1\%$ after five periods, thus this value seems like a proper $Co$ for the accurate simulation of this wave. 
 
\subsection{The effect of mesh resolution}
\begin{figure*}[ht!]
	\centering
    \includegraphics[trim=0cm 0cm 0cm 0.5cm, clip=true, scale=1]{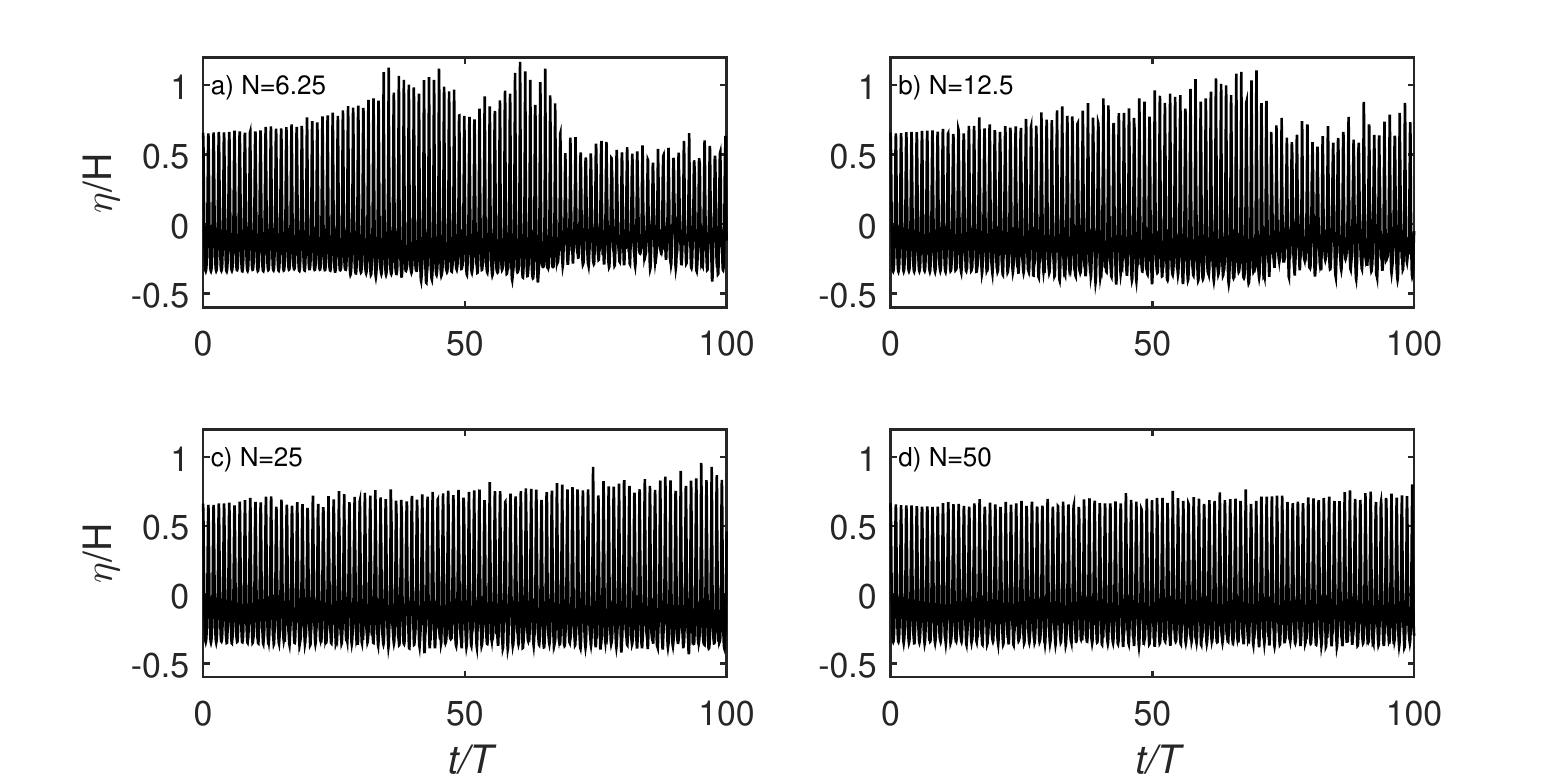}
    \cprotect\caption{Simulated surface elevation as a function of time for four different mesh resolutions (Main fixed parameters: $Co=0.15$, \verb|ddt-Euler, grad-Gauss Linear, div(rho*phi,U)-|\verb|Gauss| \verb|LimitedLinearV 1, laplacian-Gauss linear corrected| $c_\alpha =1$).}
    \label{fig:MeshEta}
 \end{figure*}

Having checked the effect of the temporal resolution, it now seems natural to check the effect of varying the spatial resolution. However, as the solution with $Co=0.5$ from the \verb|damBreak| tutorial was poor, the rest of the forthcoming analysis will be continued with $Co=0.15$, with the hope of further improving the previous results. In \cite{Jacobsenetal2012} it was noted that \verb|interFoam| performed best with cell aspect ratios, defined as $\Delta x/\Delta y$, of 1, and this ratio will be maintained throughout the analysis. In the previous cases $N=12.5$, and now three additional simulations will be performed with $N=50$, $N=25$ and $N=6.25$ respectively. Figure \ref{fig:MeshEta} shows the surface elevations as a function of time for the four different resolutions. Similar to increasing the temporal resolution (i.e.~lowering $Co$) it can be seen that increasing the number of cells per wave height greatly improves the solution when considering the ability to propagate the wave while maintaining constant form.  
 
\begin{figure}[ht]
	\centering
    \includegraphics[scale=0.6]{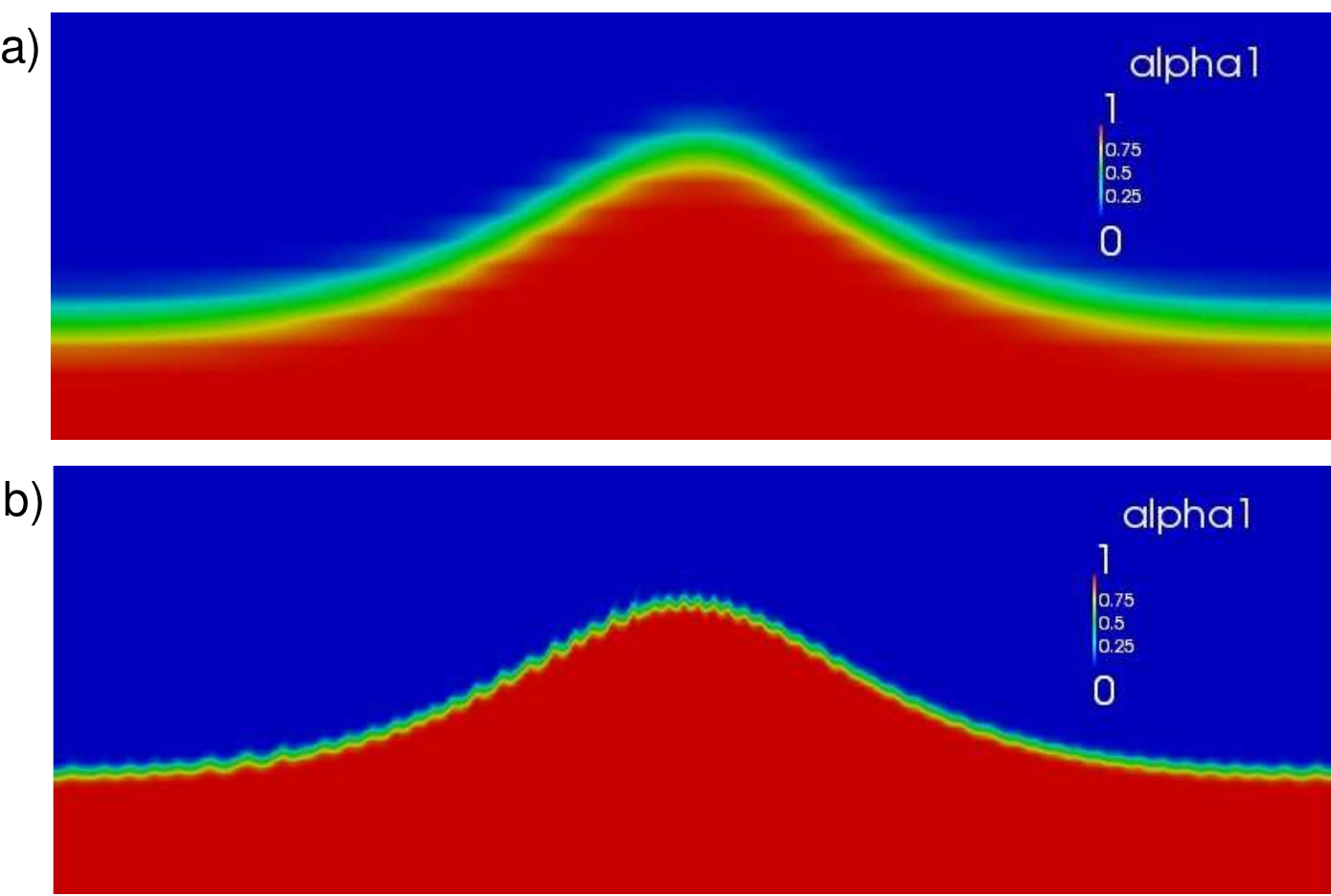}
    \cprotect\caption{Snapshot at $t=5.5 T$ for  a) $N=6.25$ and b) $N=25$  (Main fixed parameters: $Co=0.15$, \verb|ddt-Euler, grad-Gauss Linear, div(rho*phi,U)-|\verb|Gauss| \verb|LimitedLinearV 1, laplacian-Gauss linear corrected| $c_\alpha =1$).}
    \label{fig:SurfaceMesh}
 \end{figure}

Before continuing, it is also worth commenting on the shape of the air--water interface in the different resolutions, which is illustrated in Figure \ref{fig:SurfaceMesh} for $N=6.25$ and $N=25$.
As expected with $N=6.25$ the interface looks smeared and is not well captured. With $N=12.5$ (not shown here for brevity) the interface looks similar to Figure  \ref{fig:SurfacedamSetup}a, but the wave gradually steepens in time as previously explained. With $N=25$ and also $N=50$ the interface is even sharper and with $N=25$ the wave heights were also seen to increase, but somewhat slower. This is probably related to the size of the wiggles being much smaller with the finer mesh. In these cases the wiggles were not only present in the top of the crest, but also along the whole wave surface.  They also appeared at an earlier time, as seen in Figure \ref{fig:SurfaceMesh}b.  
 \begin{figure}[ht]
	\centering
    \includegraphics[scale=1]{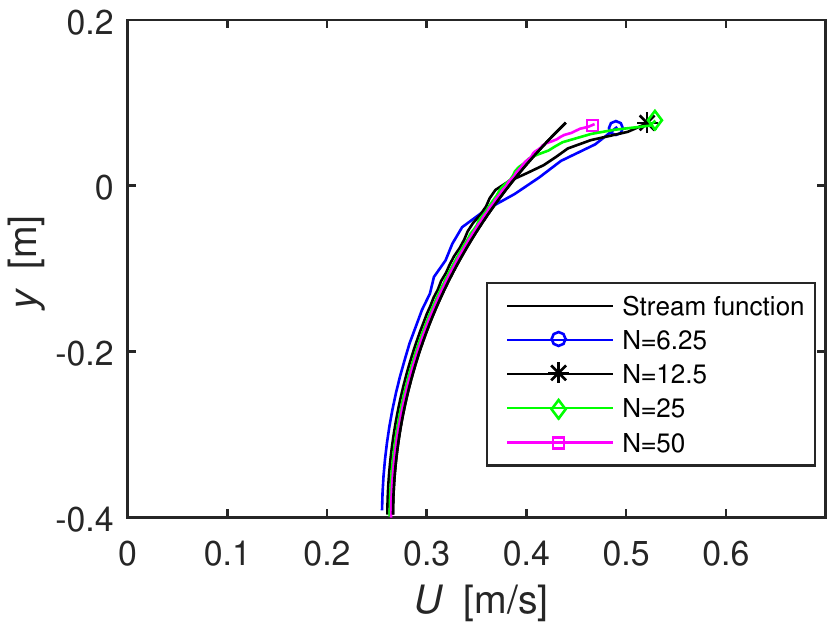}
    \cprotect\caption{Velocity distribution beneath the crest at $t=5 T$ for various mesh resolutions  (Main fixed parameters: $Co=0.15$, \verb|ddt-Euler, grad-Gauss Linear, div(rho*phi,U)-|\verb|Gauss| \verb|LimitedLinearV 1, laplacian-Gauss linear corrected| $c_\alpha =1$).}
    \label{fig:velocityMesh}
 \end{figure}
 
In Figure \ref{fig:velocityMesh} the velocity profiles beneath the crest at $t=5T$ are shown for the four different spatial resolutions together with the analytical stream function solution. In general, it can be seen that, improving the spatial resolution improves the solution. However, for the case with $N=25$ the crest velocity is as high as in the coarser resolved cases. This can be explained by the afore mentioned wiggles. At the crest of such a surface wiggle, the velocity is much higher compared to the rest of the wave. This is not seen to the same degree with $N=50$ where the surface wiggles are much smaller. When propagating the wave longer than the five periods, it was experienced that the case with $N=25$ had crest velocities closer to the analytical solution than the two coarser resolved cases. From the above results it is worth noting that increasing the spatial resolution was not able to produce as good results for the velocity profiles as increasing the temporal resolution, see Figures \ref{fig:velocityCo} and \ref{fig:velocityMesh}.  From a computational point of view decreasing $Co$ seem to be a more efficient alternative to increase accuracy, than increasing the mesh resolution. This is especially true considering that increasing the mesh resolution, will also make the time step decrease to maintain a given $Co$. However, in terms of keeping the wave height constant for the entire simulation, increasing the spatial resolution does seem to yield better results compared to simply increasing the temporal resolution.

\subsection{fvSchemes and fvSolution settings}
\label{sec:Schemes}
Thus far increasing the temporal and spatial resolution have been attempted, and  unsurprisingly, these improved the solution. For the rest of this study $Co=0.15$ and $N=12.5$ will be maintained for the sake of balancing computational costs and accuracy, and the additional effects of changing schemes and solution settings will be investigated. As quite a few schemes are available, not all results of our investigations will be shown. Our findings will be summarized and figures will be included when found to be most relevant. Later, we will combine some of the investigated schemes to improve the overall solution quality.

It has been shown that the interface between air and water in time develop wiggles, which in time grow and sometimes lead to breaking. First, the additional effects of modifying \verb|cAlpha| (with default value $c_\alpha=1$), which controls the size of the compression velocity, will be investigated. It was experienced that increasing $c_\alpha$ causes the wiggles to appear earlier and grow faster. Reducing $c_\alpha$ reduces the wiggles and at the same time causes the interface to smear out over more cells. This strongly indicates that the wiggles are caused by the numerical interface compression method. 

\begin{figure*}[ht!]
	\centering
    \includegraphics[trim=0cm 0cm 0cm 0.5cm, clip=true, scale=1]{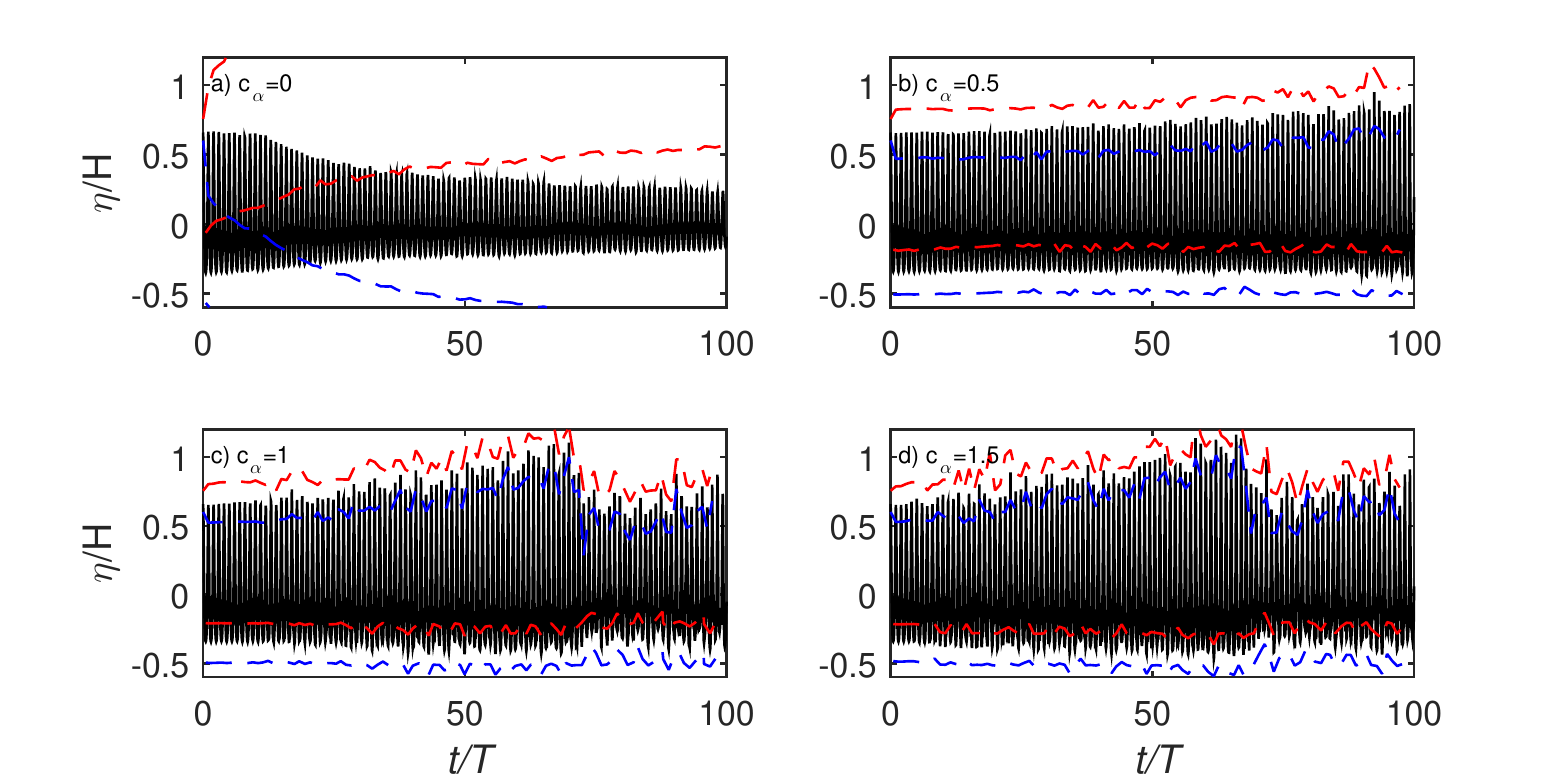}
    \cprotect\caption{Simulated surface elevations (-) as a function of time for different values of $c_\alpha$ together with the $\alpha=0.99$ and $\alpha =0.01$ contours (- -)  (Main fixed parameters: $Co=0.15$, $N=12.5$, \verb|ddt-Euler, grad-Gauss Linear, div(rho*phi,U)-|\verb|Gauss| \verb|LimitedLinearV 1, laplacian-Gauss linear corrected|).}
    \label{fig:calphaeta}
 \end{figure*}

To illustrate the effect of $c_\alpha$, the surface elevations are shown for four different values  in Figure \ref{fig:calphaeta}. In this figure, to demonstrate the effect of $c_{\alpha}$ on the interface, we also plot the $\alpha=0.99$ and $\alpha=0.01$ contours for the crest and the trough for each period.
The reduction in wave height seen in the case with $c_\alpha$=0 (Figure \ref{fig:calphaeta}a), is the effect of a very heavy diffusion of the interface. This can be seen even more clearly when looking at the $\alpha = 0.99$ and $\alpha =0.01$ contours. It can be seen that after 20 periods the 0.99 contour at the crest is actually positioned lower than the trough level and the 0.01 contour at the trough is almost at the crest level. The distance between the 0.01 contour and 0.99 contour is approximately four cells with $c_\alpha=0.5$ (Figure \ref{fig:calphaeta}b), whereas it only spans approximately three cells for $c_\alpha=1$ (Figure \ref{fig:calphaeta}c) and $c_\alpha =1.5$ (Figure \ref{fig:calphaeta}d). This shows that increasing $c_\alpha$ does compress the interface, but that the interface will span more than one cell, even with a high value of $c_\alpha$.

\begin{figure}[ht!]
	\centering
    \includegraphics[scale=1]{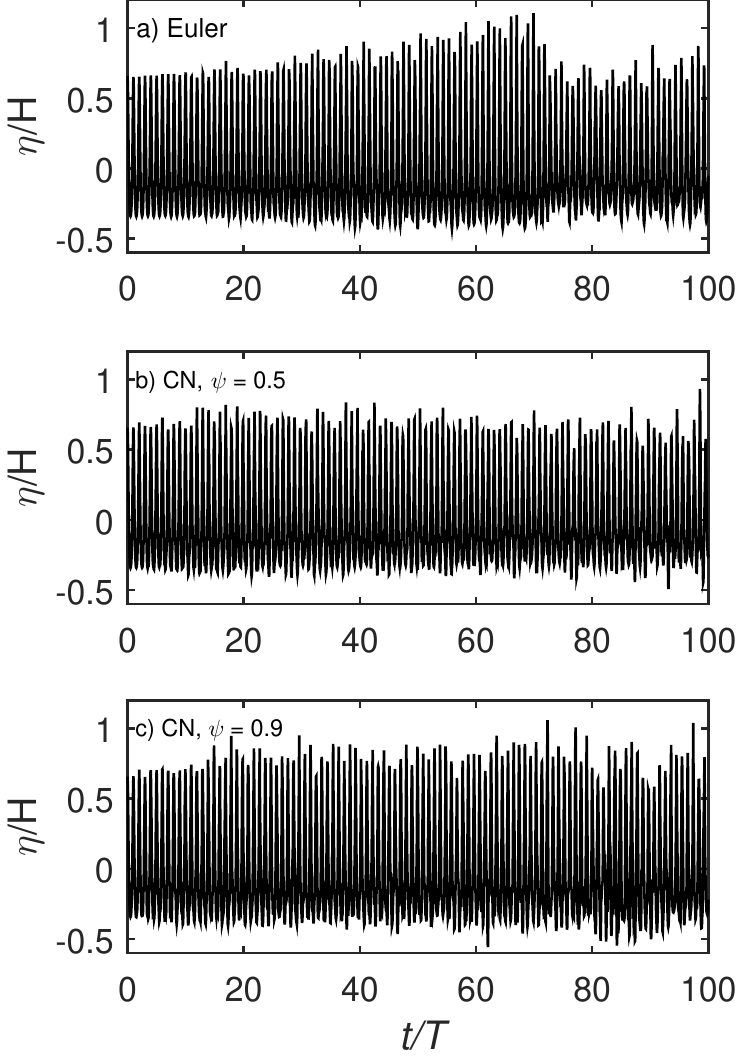}
    \cprotect\caption{Simulated surface elevation as a function of time for different time discretization schemes (Main fixed parameters: $Co=0.15$, $N=12.5$, \verb|grad-Gauss Linear, div(rho*phi,U)-|\verb|Gauss| \verb|LimitedLinearV 1, laplacian-Gauss linear corrected|, $c_\alpha=1$).}
    \label{fig:DtEta}
 \end{figure}

In addition to the $c_\alpha$ value, various other settings affect the size and behaviour of the wiggles, and in the following $c_\alpha=1$ will be maintained, for the sake of comparison. The effect of the time discretization scheme on the surface elevations is shown in Figure \ref{fig:DtEta}.  Changing the time discretization scheme from \verb|Euler| (first order) to CN (second order) exacerbates the wiggle feature, causing them to develop earlier and extend throughout the surface. Contrary to results utilizing the \verb|Euler| scheme, the wiggles do not cause the wave to steepen to the same extent. The wiggles grow in size, but they often break on top of the wave before merging, and therefore the wave does not steepen as much as with the \verb|Euler| scheme. 
It is believed that the wiggle feature is more pronounced with the CN scheme simply because the scheme is less diffusive than the \verb|Euler| scheme. The artificial compression term, as just shown, adds some erratic behaviour to the interface, and this is diffused by numerical damping when using the \verb|Euler| scheme, but less so when using CN. 

The reduction or complete removal of wiggle formations is also seen utilizing other more diffusive schemes, e.g.~when using the upwind scheme for the convection of the $\alpha$ field or using the upwind scheme for the convection of momentum. In the case of utilizing the upwind scheme for the convection of the $\alpha$ field the solution is very similar to that seen when setting $c_\alpha =0$ (Figure \ref{fig:calphaeta}a), with the interface experiencing heavy diffusion and the resulting wave height decaying rapidly. Utilizing an upwind scheme for the convection of momentum also causes the wave height to decay, but at a much slower rate, and is not accompanied by the same degree of interface diffusion. However, utilizing a pure upwind scheme is generally not recommended due to excessive smearing of the solution.

Thus far it has been shown that $c_\alpha$ and the time discretization scheme have a significant impact on the surface elevation and interface. However, regarding the velocity profile beneath the crest (not shown here for brevity), the impact is very small, except for the case with $c_\alpha=0$, which made made the velocities throughout the water column beneath the crest too low. This is probably due to heavy diffusion of the interface (see Figure \ref{fig:calphaeta}a).

\begin{figure*}[ht!]
	\centering
    \includegraphics[trim=0cm 0cm 0cm 0.5cm, clip=true, scale=1]{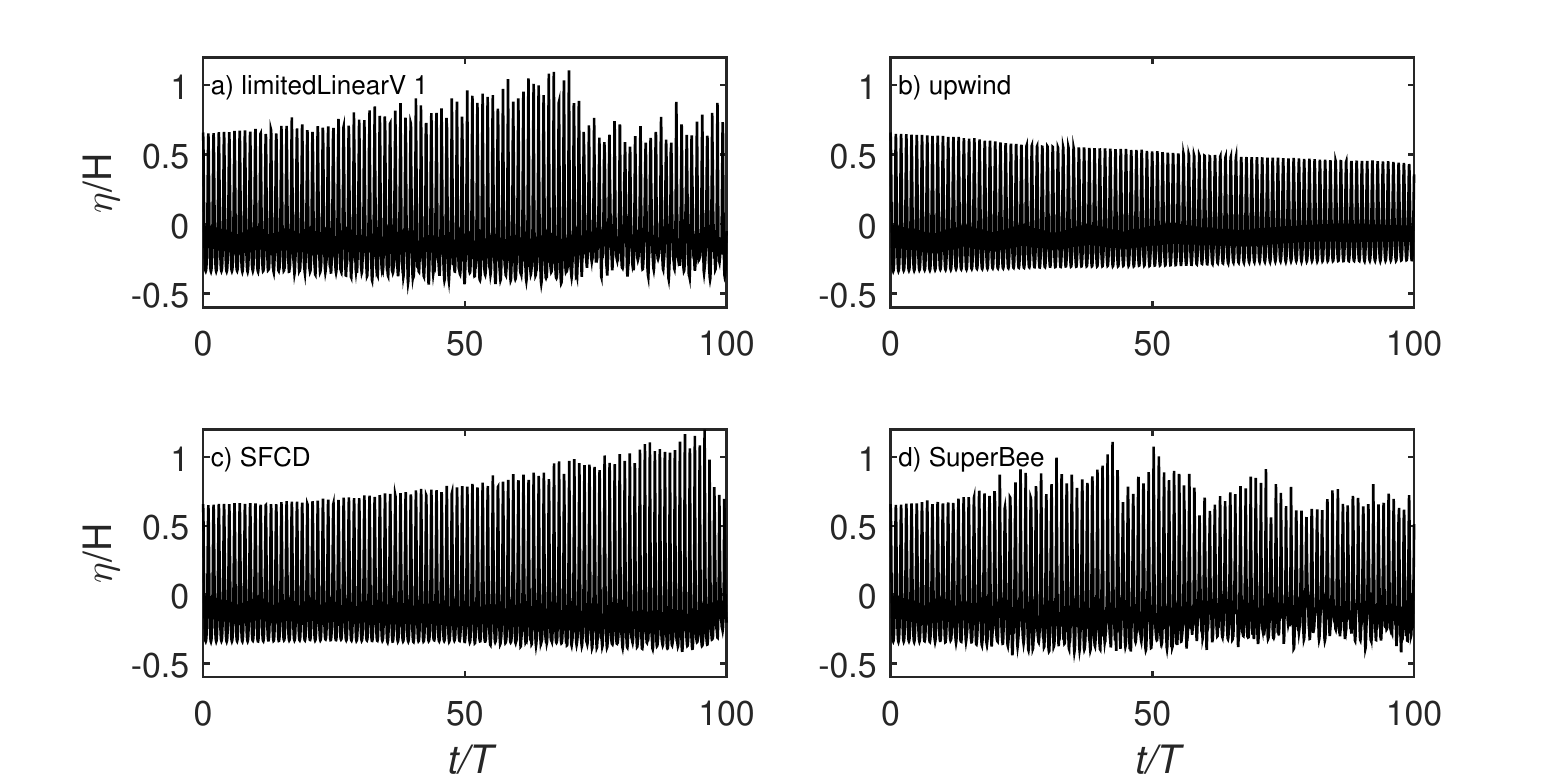}
    \cprotect\caption{Simulated surface elevation as a function of time for different convection schemes (Main fixed parameters: $Co=0.15$, $N=12.5$, \verb|ddt-Euler, grad-Gauss Linear, laplacian-Gauss linear| \verb|corrected|, $c_\alpha=1$).}
    \label{fig:Choseneta}
 \end{figure*}

As mentioned, the wiggles can be limited by choosing more diffusive schemes, but it still needs to be determined how these schemes affect the general propagation of the wave and the underlying velocity profile. 
Figure \ref{fig:Choseneta} shows the surface elevation for four different convection schemes (\verb|div(rho*phi,U)|), and the influence of the choice on convection scheme is readily apparent. The most diffusive among the four schemes, the \verb|upwind| scheme, makes the wave decay in a quite stable fashion (Figure \ref{fig:Choseneta}b). The \verb|SFCD| scheme (Figure \ref{fig:Choseneta}c) is slightly more diffusive than the \verb|limitedLinearV 1| scheme (Figure \ref{fig:Choseneta}a), and is seen to limit the growth in the wave height. The wave height still increases as time progresses but the increase is delayed and the simulation is less erratic. The fourth scheme is the \verb|SuperBee| scheme (Figure \ref{fig:Choseneta}d). This scheme is also within the TVD family, but it is much more erratic, and almost immediately the wave heights start to increase. 

\begin{figure}[ht!]
	\centering
    \includegraphics[scale=1]{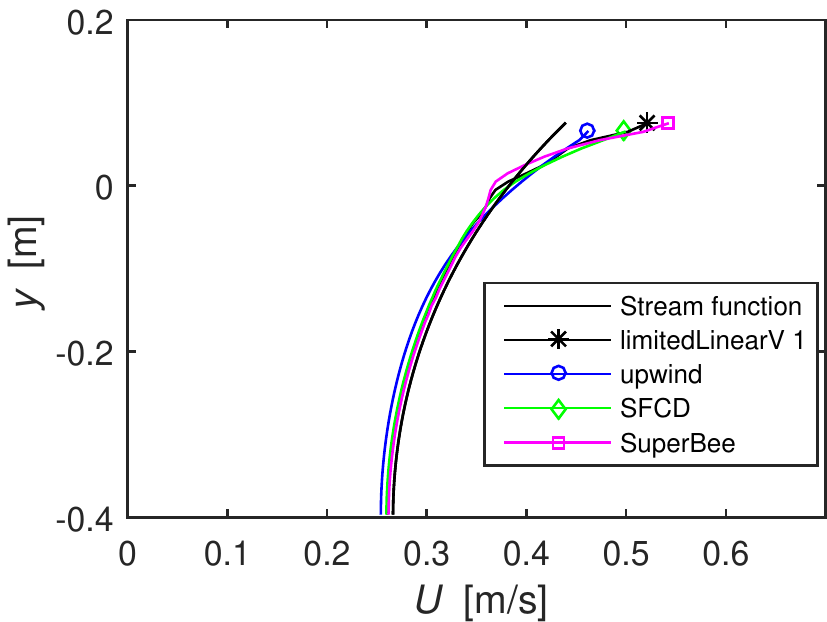}
    \cprotect\caption{Velocity distribution beneath the crest at $t=5T$ for various convection schemes (Main fixed parameters: $Co=0.15$, $N=12.5$, \verb|ddt-Euler, grad-Gauss Linear, laplacian-Gauss linear| \verb|corrected|, $c_\alpha=1$).}
    \label{fig:velocityDiv}
 \end{figure}

The velocity profiles beneath the crest for the four convection schemes are likewise shown at $t=5T$ in Figure \ref{fig:velocityDiv}, and once again the importance of the convection scheme is quite clear. The \verb|upwind| scheme limits the error in the velocity at the top crest whereas it underestimates the velocity closer to the bed. The \verb|SFCD| scheme behaves slightly better than the \verb|limitedLinearV 1| scheme, and the \verb|SuperBee| scheme performs the worst. When propagating further the \verb|SuperBee| scheme has oscillations in the velocity profile beneath the crest, which can also be seen to a smaller degree in Figure \ref{fig:velocityDiv}. 

A range of other convection schemes have also been attempted. None of them, however, show significantly different results than those shown here, which have been selected to demonstrate the effect of convection scheme diffusivity on the propagation of the wave and velocity profile beneath the crest. While the convection schemes have been shown to have a great effect on both the ability to maintain a constant wave height, limit the wiggle feature in the interface and predict the velocity profile, it is not directly evident which scheme performs the best overall. The \verb|upwind| scheme limits the error in the crest velocity the most, which would be beneficial when e.g. doing loads on structures, but due to the diffusivity of the scheme might not be able to capture e.g. vortex shedding around such a structure. The \verb|SFCD| scheme improves the ability to maintain a constant wave height and limits the growth in the crest velocity compared to the \verb|limitedLinearV 1| scheme from the \verb|damBreak| tutorial, but the crest velocity is still severely overestimated.

We will now turn our attention to the gradient (\verb|grad|) schemes. These effects (relative to the default \verb|Gauss Linear| scheme in Figures \ref{fig:CoElevation}c and \ref{fig:velocityCo}) on the wave propagation and velocity profile will be described, but for brevity no additional figures will be included.
The fourth-order scheme (\verb|fourth|) improves the propagation and delays the increase in wave heights, similar to the behaviour seen with the \verb|SFCD| convective scheme (Figure \ref{fig:Choseneta}c), which is more diffusive than the standard \verb|limitedLinearV 1| scheme.  The \verb|fourth| scheme is however not more diffusive than the \verb|Gauss Linear| scheme, and the delayed increase in wave height is probably due to the scheme having higher-order accuracy. The velocity profile beneath the crest, on the other hand, is not improved relative to the \verb|Gauss Linear| scheme (Figure \ref{fig:velocityCo}, $Co$=0.15).  The \verb|faceMDLimited Gauss Linear 1| gradient scheme has also been tested, and behaves very similar to the \verb|upwind| convection scheme (Figure \ref{fig:Choseneta}b), in the sense that the wave heights decrease with time. The reason for this is probably that the gradient limiter, coupled with the \verb|limitedLinearV 1| convection scheme, effectively makes the convection scheme an upwind scheme. With respect to the velocities the \verb|faceMDLimited| gradient scheme produced a velocity profile very similar to that from the \verb|upwind| scheme (Figure \ref{fig:velocityDiv}). That the limited gradient scheme can produce results similar to the \verb|upwind| convection scheme was also observed by	\cite{LiuHinrichsen2014}, who studied the effect of convection and gradient schemes on bubbling fluidized beds using \verb|OpenFOAM|. 

\begin{figure}[ht!]
	\centering
    \includegraphics[scale=1]{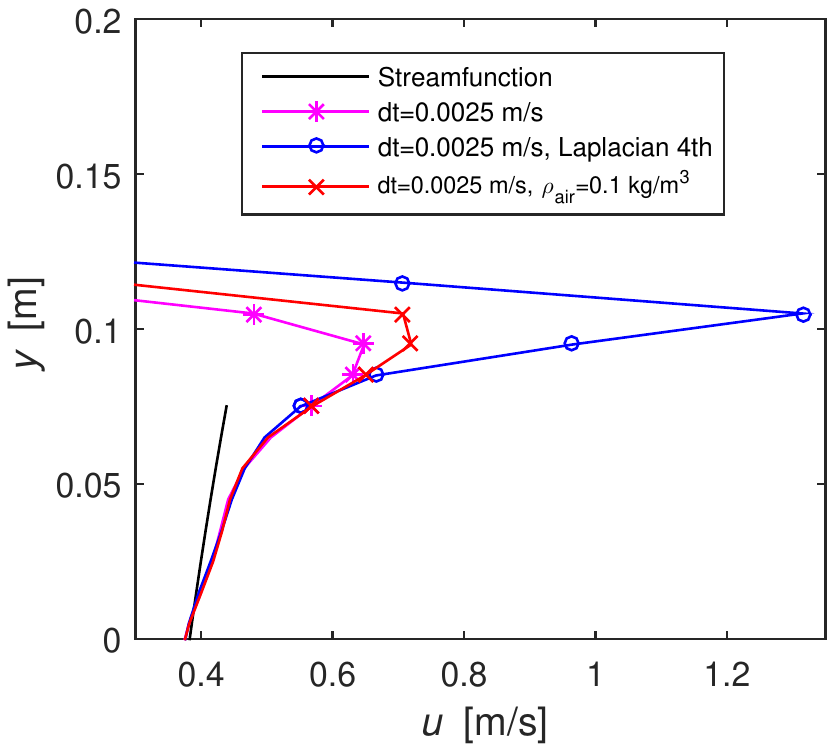}
    \cprotect\caption{Velocity distribution beneath the crest at $t=5t$ with a fixed time step utilizing the standard setup as well as 4th order Laplacian and $\rho_{\text{air}}=0.1$ kg/m$^3$. Full lines represent the velocities in pure water and the lines with symbols represent the velocities in the air or mixture cells (Main fixed parameters: $N=12.5$, \verb|ddt-Euler, grad-Gauss Linear, div(rho*phi,U)-|\verb|Gauss| \verb|LimitedLinearV 1|, $c_\alpha=1$)}
    \label{fig:velocityFixedDt}
 \end{figure}

We will now describe how changing the Laplacian scheme effects the solution, relative to the default setting (\verb|Gauss linear corrected|). As previously mentioned the Laplacian scheme requires keywords for both \verb|interpolation| and \verb|snGrad|, but the inputs for the stand alone \verb|interpolation| and \verb|snGrad| schemes are not changed.
For the Laplacian scheme, combining the \verb|Gauss linear| interpolation with the \verb|fourth| \verb|snGrad| scheme, resulting in the Laplacian scheme \verb|Gauss Linear fourth|, gave improved results, both in terms of the ability to maintain constant wave heights and in terms of the velocity profile beneath the crest. However switching to the fourth-order scheme (\verb|fourth|), resulted in very high spurious velocities in the air region above the wave, and hence (due to the $Co$-controlled time step) leads to reductions in the time steps used during the simulation. In this way changing to a fourth-order \verb|snGrad| schemes in the Laplacian is effectively similar to lowering $Co$.  To check whether the fourth-order \verb|snGrad| scheme in the Laplacian really improved the solution, or if it is merely a result of a reduced time step, two additional simulations, now utilizing a fixed time step dt=0.0025 m/s, have been performed, with both \verb|corrected| and \verb|fourth| \verb|snGrad| scheme in the Laplacian.  The resulting velocity profiles at $t=5T$, together with the result from a simulation with $\rho_{air}=0.1$ kg/m$^3$ (also utilizing the same fixed time step), are shown in Figure \ref{fig:velocityFixedDt}.
The three simulations show similar results in the water phase, but rather different velocities in the air phase. These results indicate that, while being an un-physical and undesirable phenomenon, the spurious velocities in the air do not seem to effect the wave significantly. The case with a fourth-order \verb|snGrad| scheme had approximately twice as high air velocities as the standard set up, but similar (actually slightly lower) crest velocities. The case with lower density also has higher air velocities, but very similar water velocities to the standard case.  To summarize: Even though the fourth-order Laplacian scheme is able to produce better wave kinematics, caution must be taken as it produces large spurious velocities. These will, utilizing a variable time step, lead to very low time steps. Alternatively, a fixed time step may result in an unstable Courant number.

Before conducting the present study it was expected that the discretization schemes would have an effect on the solution, but it was also expected that in particular the choice of iterative solvers for the pressure would not have an effect, at least if the tolerances were sufficiently low.  It turns out, however, that the iterative solver settings in \verb|fvSolution| also affect the wave propagation. For the pressure equations (\verb|pcorr|, \verb|pd| and \verb|pdFinal|)  switching from \verb|PCG| to \verb|GAMG| made the simulations more erratic as the wave broke much earlier (however the simulation time was much lower), whereas switching to a smooth solver (\verb|smoothSolver|) did not affect the quality of the solution, but took much longer time. It was also attempted to lower the tolerance by a factor 1000 on both the pressure and the velocity, but hardly any difference in the solution was seen. For the controls of the solution algorithm increasing the number of alpha correctors, \verb|nAlphaCorr|, as well as alpha subcycles, \verb|nAlphaSubCycles|, improved, though not dramatically, the propagation of the wave in terms of it maintaining its' shape, whereas increasing the number of correctors, \verb|nCorrectors| did not change anything. Increasing the number of outer correctors, \verb|nOuterCorrectors| (nOCorr), effectively making it into the PIMPLE algorithm, surprisingly made the wave height decrease very rapidly. This behaviour was also seen in \cite{Weber2016} and will be investigated further in the forthcoming section.

The choice of iterative solvers could also potentially effect the velocity profile. The \verb|GAMG| solver produced much higher crest velocities (close to that seen with $Co=0.5$ in Figure \ref{fig:velocitydamSetup}). The \verb|SmoothSolver|, which was a lot slower, produced an almost identical velocity profile to the \verb|PCG| solver (Figure \ref{fig:velocityCo}, $Co=0.15$). Lowering the tolerances by a factor 1000 had almost no effect on the surface elevation, and the effect on the velocity profile was also negligible. Changing the number of $\alpha$ subcycles (\verb|nAlphaSubCycles|), $\alpha$ correctors (\verb|nAlphaCorr|) and number of correctors (\verb|nCorrectors|) did not influence the crest velocity in any significant way, and raising the number of $\alpha$ correctors actually worsened the result closer to the bed. 

It has now been shown that the discretization schemes and solution procedures have a potentially large impact in the solution, both in terms of the wave height and velocity profile, as well as the wiggles in the interface and the spurious air velocities. Using more diffusive schemes than the base setup from the \verb|damBreak| tutorial has been shown to limit or remove the growth of the wiggles, limit the overestimation of the crest velocity, and also limit the growth of the wave heights. However, the more diffusive schemes were seen to smear the interface, and could potentially be more inaccurate for other situations.

\subsection{Combined schemes}
It would be ideal to achieve a setup capable of propagating a wave for 100 periods, while keeping a relatively large time step and at the same time maintaining both its shape and the correct velocities. Changing one single scheme has not achieved that. It was however shown that adding some diffusion in some of the schemes could mitigate both the increase in wave height as well as the increased near-crest velocities. 

To test whether a combination of schemes can improve the solution further, the \verb|upwind| scheme on the convection of momentum, which was actually seen to cause the wave to decay (Figure \ref{fig:Choseneta}b), will be combined with the slightly less diffusive blended CN scheme (Figure \ref{fig:DtEta}c). It is also attempted to increase the artificial compression, by increasing $c_\alpha$ while picking a more diffusive scheme for the gradient, namely \verb|faceMDLimited| which also caused the wave height to decrease. Finally, the outer correctors  are increased to two and combined with the blended CN scheme, together with the \verb|SFCD| scheme for the momentum flux.

\begin{figure*}[ht!]
	\centering
    \includegraphics[trim=0cm 0cm 0cm 0.5cm, clip=true, scale=1]{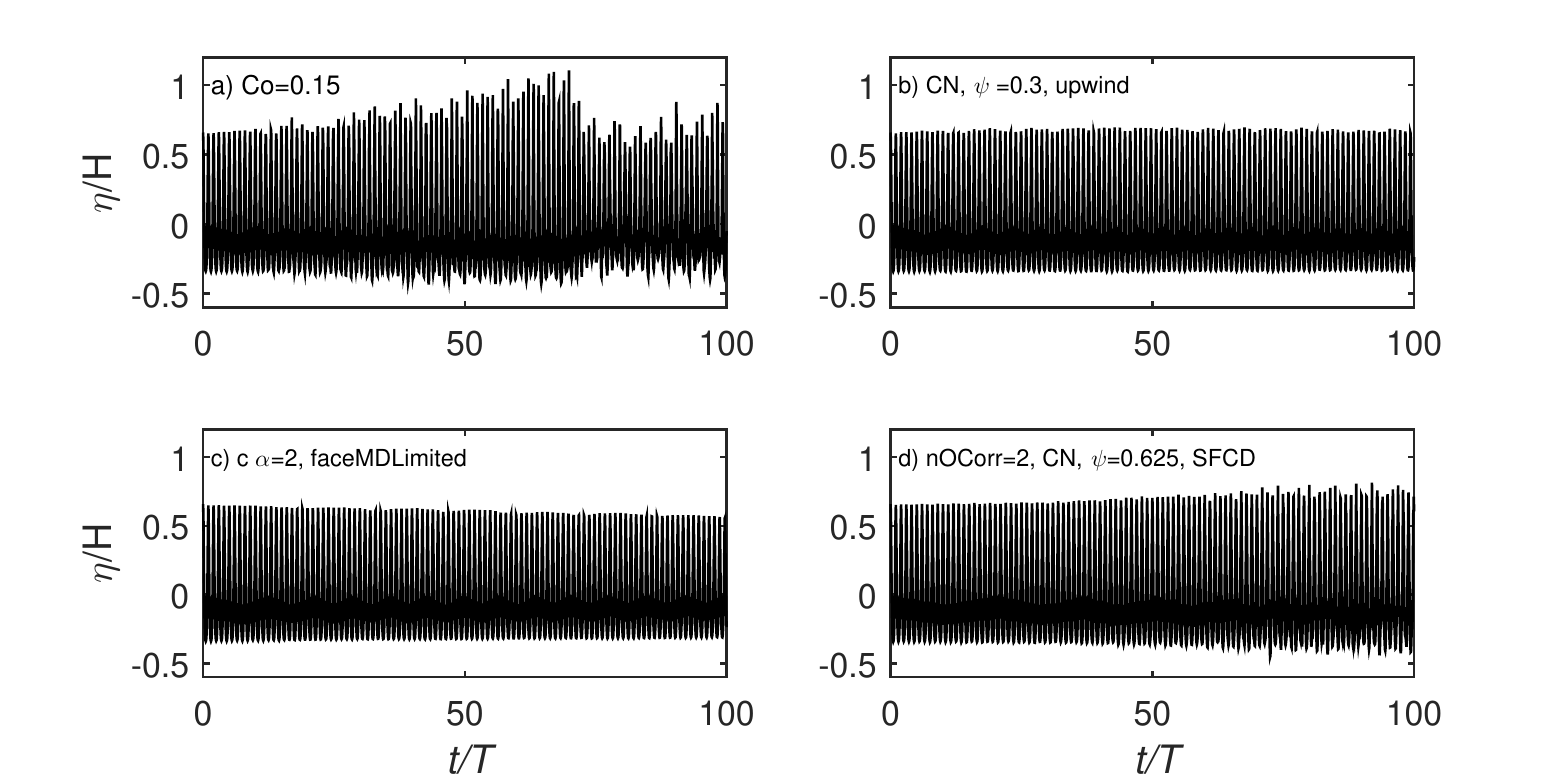}
    \cprotect\caption{Simulated surface elevation as a function of time for different schemes (Main fixed parameters: $Co=0.15$, $N=12.5$, \verb|laplacian-Gauss linear corrected|).}
    \label{fig:combinedEta1}
 \end{figure*}

The surface elevations for three such combinations are seen in Figure \ref{fig:combinedEta1}b--d.  Here it can be seen that by combining the diffusive \verb|upwind| scheme for the convection of momentum and shifting from the more diffusive \verb|Euler| scheme to a less diffusive CN scheme (Figure \ref{fig:combinedEta1}b) can maintain he wave height for the entire 100 periods. The same can be done by increasing the compression factor $c_\alpha$ while maintaining a more diffusive gradient scheme (Figure \ref{fig:combinedEta1}c, although in this case the wave heights actually decayed a bit), and also by increasing the number of outer correctors together with the CN scheme (Figure \ref{fig:combinedEta1}d). The latter results in slightly more variations in the wave height, but also utilized a much higher blending value in the CN scheme, which can cause oscillations in the solution and, as previously shown, excite wiggles in the free surface.
All three cases show a great improvement compared to the original default case, repeated as Figure \ref{fig:combinedEta1}a to ease comparison.
It should also be stated that the balance obtained for the case with the outer correctors is particularly delicate. First it was attempted to run with two outer correctors and a blended CN scheme, while maintaining the \verb|limitedLinearV 1| scheme on the momentum flux. This however caused wiggles in the interface, as also previously described, and therefore the \verb|SFCD| scheme was chosen to counteract the wiggles. The wiggles were not removed altogether with the \verb|SFCD| scheme, but their presence was significantly delayed.  Further, the best result was obtained with CN, $\psi=0.625$, but lowering the blending factor to $\psi=0.6$ made the wave height decrease slightly over the 100 periods, and raising it to $\psi=0.65$ made it increase slightly and caused more wiggles.

\begin{figure}[ht!]
	\centering
    \includegraphics[trim=0cm 0cm 0cm 0.5cm, clip=true, scale=1]{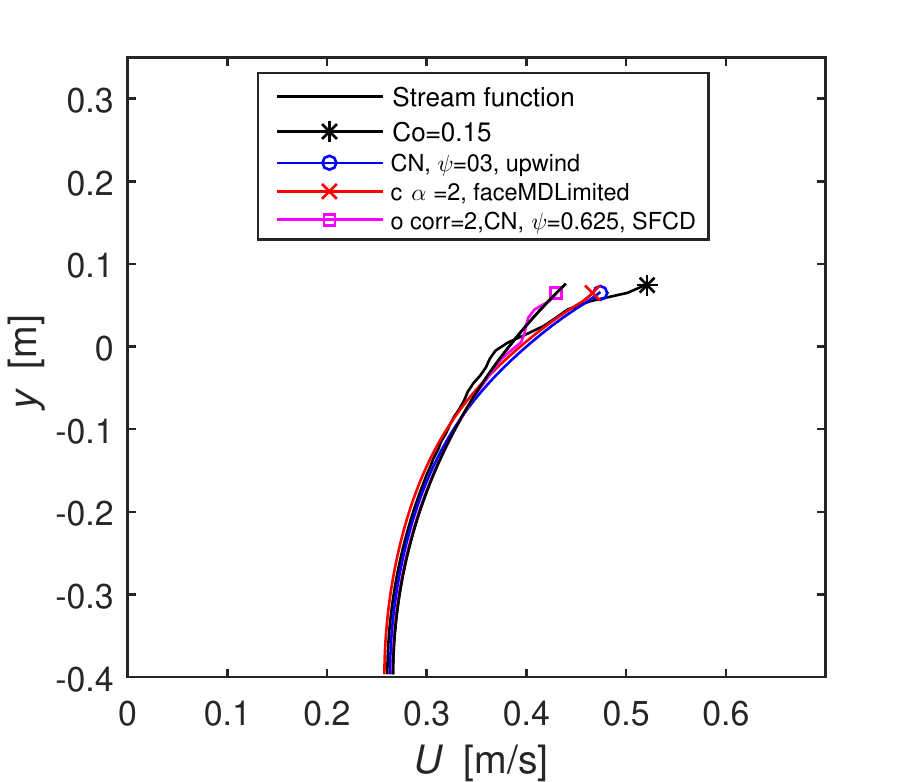}
     \cprotect\caption{Velocity distribution beneath the crest at $t=5T$ for various combined schemes (Main fixed parameters: $Co=0.15$, $N=12.5$, \verb|laplacian-Gauss linear corrected|).}
    \label{fig:combinedVelocity1}
 \end{figure}

The resulting velocity profiles beneath the crest at $t=5T$ for the three cases shown in Figure \ref{fig:combinedEta1}b-d are shown in Figure \ref{fig:combinedVelocity1}, together with the velocity profile obtained utilizing the base settings.
Here it is evident that all three combinations give lower velocities in the crest than the standard setting. However the standard setup shows a slightly better comparison with the analytical result closer to the bottom than the case utilizing \verb|upwind| for the momentum flux together with CN as well as the case utilizing $c_\alpha=2$ together with the \verb|faceMDLimited| gradient scheme. The final combination, utilizing two outer correctors together with a blended CN scheme and a \verb|SFCD| scheme shows a significantly better result, and is very similar to the analytical profile. It can be seen that there are small odd oscillations in the profile of this case, and these oscillations actually become larger as the wave propagates. Nevertheless, this significant improvement is achieved with minimal increase in computational expense, especially compared to the results obtained utilizing the settings from the \verb|damBreak| tutorial.  The improvement in the velocity profile with the outer correctors is interpreted as the outer correctors ensuring a better coupling between velocity, pressure and the free-surface.

It has now been shown that it is possible to achieve a "diffusive balance" in the schemes, that enables \verb|interFoam| to progress the wave while maintaining its shape. The same diffusive balance is also shown to limit, but (except for the case utilizing outer correctors) not eliminate, the overestimation of the velocity in the crest. This diffusive balance is, however, not universal. What seems a proper amount of diffusion in the case of $Co=0.15$ is not so with a lower $Co$ where the error in velocity of the crest is much smaller, and more diffusive schemes would actually worsen the solution. Also, what gives the best balance for this wave, might not give the best balance for a wave with another shape, but the present study reveals a generic strategy that can be fine tuned for individual cases. Interestingly, this implies that for variable depth problems, where waves would not maintain a constant form, there may not be a globally optimal combination. Nevertheless, it is still hoped that better-than-default accuracy can be achieved with the combinations suggested herein. 

\subsection{Summary of experience using interFoam}

To summarize our experience using \verb|interFoam| from this section: The safest way to get a good and stable solution is by using a small Courant number. If the time step is low enough, \verb|interFoam| is capable of producing quite good results. However, due to limited time or computational resources, this solution may often not be realistic in practice. 

If wishing to use larger time steps, alternatively, it is advised to try to obtain a diffusive balance. The best choice can then be determined on a case by case basis, though it is hoped that the examples utilized above may be a good starting point for more general situations. If looking to simulate e.g.~wave breaking, the incoming waves could first be simulated in a cyclic domain, as done herein, prior to doing the actual larger-scale simulation. In this smaller simulation, the proper balance between, diffusivity, time step, computational expense and solution accuracy could be determined, before doing more advanced simulations. This should help ensure that reasonable accuracy in the initial propagation is maintained, which is important as this will affect the initial breaking point and hence the subsequent surf zone processes.

\begin{figure}[t]
	\centering
    \includegraphics[scale=1]{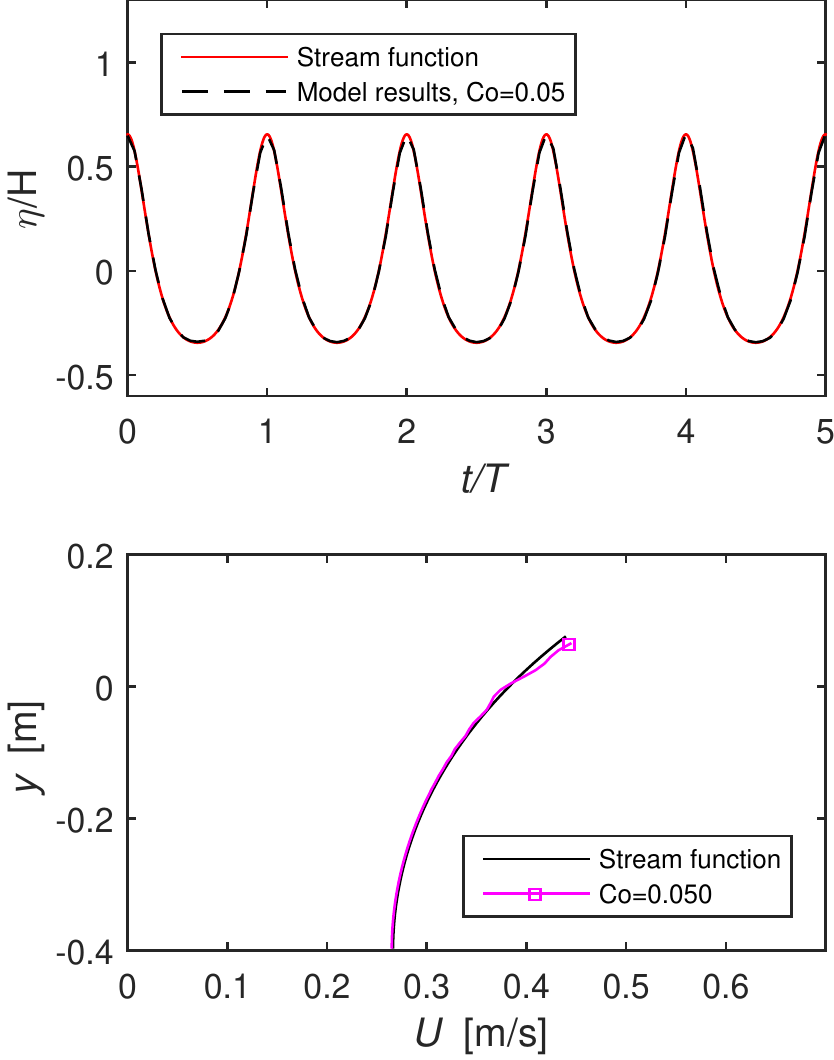}
    \cprotect\caption{Surface elevations and velocity distribution beneath the crest at $t=5T$ for $Co=0.05$ (Main fixed parameters: $N=12.5$, \verb|ddt-Euler, grad-Gauss Linear, div(rho*phi,U)-| \verb|Gauss LimitedLinearV 1, laplacian-Gauss linear corrected|, $c_\alpha =1$). }
    \label{fig:CoStrength}
 \end{figure}

The present results have focused on a rather demanding task of simulating long-time CFD wave propagation over 100 periods, though the problem with the overestimation of crest velocities show up much earlier (see again Figure \ref{fig:velocitydamSetup}).  To underline that \verb|interFoam| is capable of producing a good result for most practical applications involving shorter propagation horizons, without having to resort to a diffusive balance strategy, Figure \ref{fig:CoStrength} shows the surface elevations for the first five periods, as well as the velocity profile beneath the crest at $t=5T$ using a small $Co=0.05$. Here a good match with the analytical stream function solution is achieved. A similar improvement in the prediction of the crest velocities, with reduction of Courant number, were shown in \cite{Roenbyetal2017a}, and this thus seems to be a robust and generally viable strategy.  

\section{interFoam coupled with isoAdvector: interFlow}

One of the problems with \verb|interFoam| is that the surface gets smeared over several cells, as demonstrated in Section \ref{sec:Schemes}. This is mitigated by the artificial compression term, which makes the surface sharper, but (as shown herein, Figure \ref{fig:calphaeta}) also produces some undesired effects. In this section we will finally test the results using \verb|interFoam| coupled with the \verb|isoAdvector| algorithm, recently developed by \cite{Roenbyetal2016}, which is also available in the newest version of \verb|OpenFOAM| (\verb|OpenFOAM-v1706|). The \verb|isoAdvector| version in \verb|OpenFOAM-v1706| has a slightly different implementation of the outer correctors than the version used in the present study, see \cite{Roenbyetal2017} for details. 
With \verb|isoAdvector| the equation for $\alpha$ \eqref{eq:alpha1} is not solved directly. Instead the surface is identified by an iso-line, similar to those shown for $\alpha=0.99$ and $\alpha=0.01$ in Figure \ref{fig:calphaeta}. After identifying the exact position of the surface, it is then advected in a geometric manner. For more details on the implementation of \verb|isoAdvector| the reader is referred to \cite{Roenbyetal2016}.

The new \verb|isoAdvector| algorithm, coupled with \verb|interFoam| will for the remainder of this study be named \verb|interFlow|. As a first case,  \verb|interFlow| and \verb|interFoam| will be compared for the previously well-tested case with the \verb|damBreak| settings and $Co=0.15$. It should be stated however, that \verb|interFlow| was not able to propagate the wave with the settings used in \verb|interFoam|. The tolerances on $p^*$ (\verb|pd|) needed to be reduced by a factor 100 and the tolerances on $U$ (\verb|U|) by a factor 10. Comparing the performance of the two is, however, still justified as \verb|interFlow| actually, even with the decreased tolerances, performed the simulation slightly faster than \verb|interFoam|. Moreover, the simulations with \verb|interFoam| did not improve when lowering the tolerances with a factor 1000 as shown in Section \ref{sec:Schemes}. The speed-up in computational time was not due to larger time steps, but rather to the algorithm moving the free surface faster.

\begin{figure*}[ht!]
	\centering
    \includegraphics[trim=0cm 0cm 0cm 0.2cm, clip=true, scale=1]{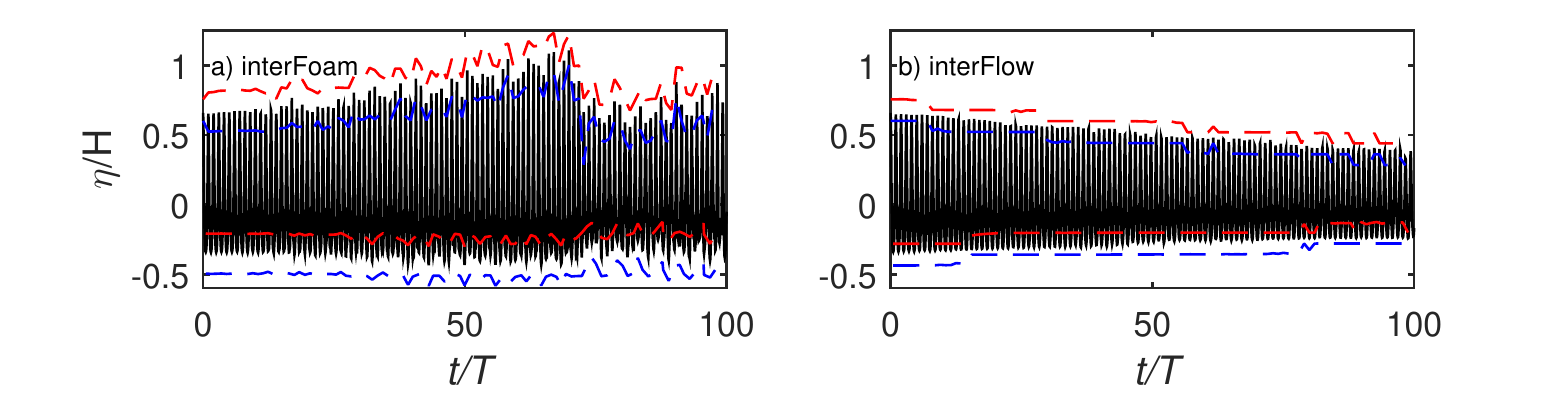}
    \cprotect\caption{Simulated surface elevations (-) as a function of time utilizing interFoam and interFlow together with the $\alpha=0.99$ and $\alpha=0.01$ (- -) contours (Main fixed parameters: $Co=0.15$, $N=12.5$, \verb|ddt-Euler, grad-Gauss Linear, div(rho*phi,U)-| \verb|Gauss LimitedLinearV 1, laplacian-Gauss linear corrected|).}
    \label{fig:FlowFoameta}
 \end{figure*}

Figure \ref{fig:FlowFoameta} shows the surface elevations obtained utilizing the two different solvers. It is quite noticeable that, while with \verb|interFoam| the wave heights start to increase, with \verb|interFlow| the wave heights decrease mildly. Also shown are the contours for $\alpha =0.99$ and $\alpha=0.01$ for the crest and trough for each period. Here it can be seen that the two contours are substantially closer with \verb|interFlow|. They are constantly separated by less than two cell heights meaning that there is actually only one interface cell in the vertical direction. This is a substantial improvement of the surface representation compared to \verb|interFoam|. Since equation \eqref{eq:alpha1num} is not solved, there is no artificial compression term, and the interface wiggles previously observed are gone altogether. This is likewise a desirable improvement. The artificial compression term has been shown to have undesired effects, as it cause wiggles in the interface, in the simple propagation of a stream-function wave over sufficiently long propagation times. How these wiggles might behave in more complex situation like e.g.~wave breaking is an open question, but one can imagine a greater effect in such a more chaotic situation. 
 
\begin{figure}[ht]
	\centering
    \includegraphics[scale=1]{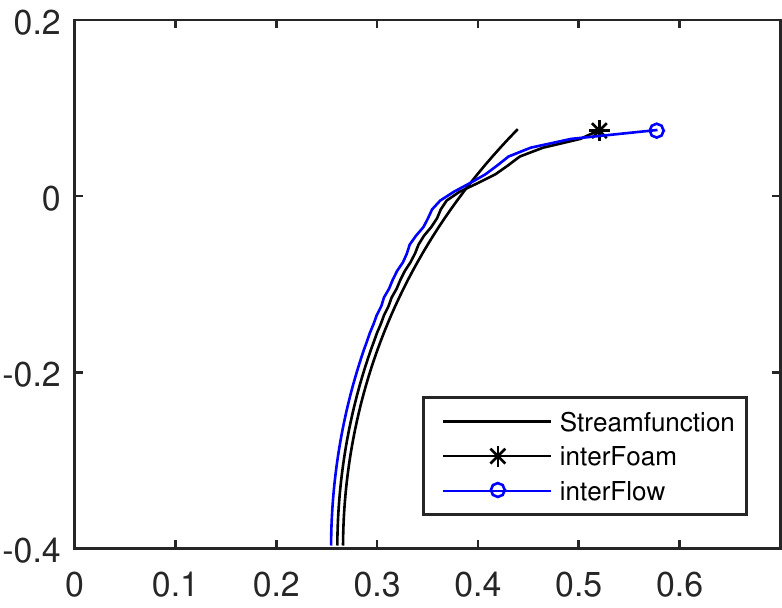}
    \cprotect\caption{Velocity distribution beneath the crest at $t=5T$ utilizing interFoam and interFlow (Main fixed parameters: $Co=0.15$, $N=12.5$, \verb|ddt-Euler, grad-Gauss Linear, div(rho*phi,U)-| \verb|Gauss LimitedLinearV 1, laplacian-Gauss linear corrected|).}
    \label{fig:FlowFoamVelocity}
 \end{figure}

In Figure \ref{fig:FlowFoamVelocity} the velocity profile beneath the crest at $t=5T$ is shown utilizing both \verb|interFoam| and \verb|interFlow|. Here it is quite clear that \verb|interFlow|, with the current settings is not improving the velocity profile. The crest velocity is slightly larger than the \verb|interFoam| solution, and closer to the bed, the velocity is underestimated. This underestimation of velocity is probably due to the decrease in wave height.
That \verb|interFlow| gets an even larger error in the velocity in the top of the crest is probably due to the sharper interface, creating larger gradients, and any imbalance in the momentum equation near the interface may then be increased.

\begin{figure}[ht]
	\centering
    \includegraphics[scale=1]{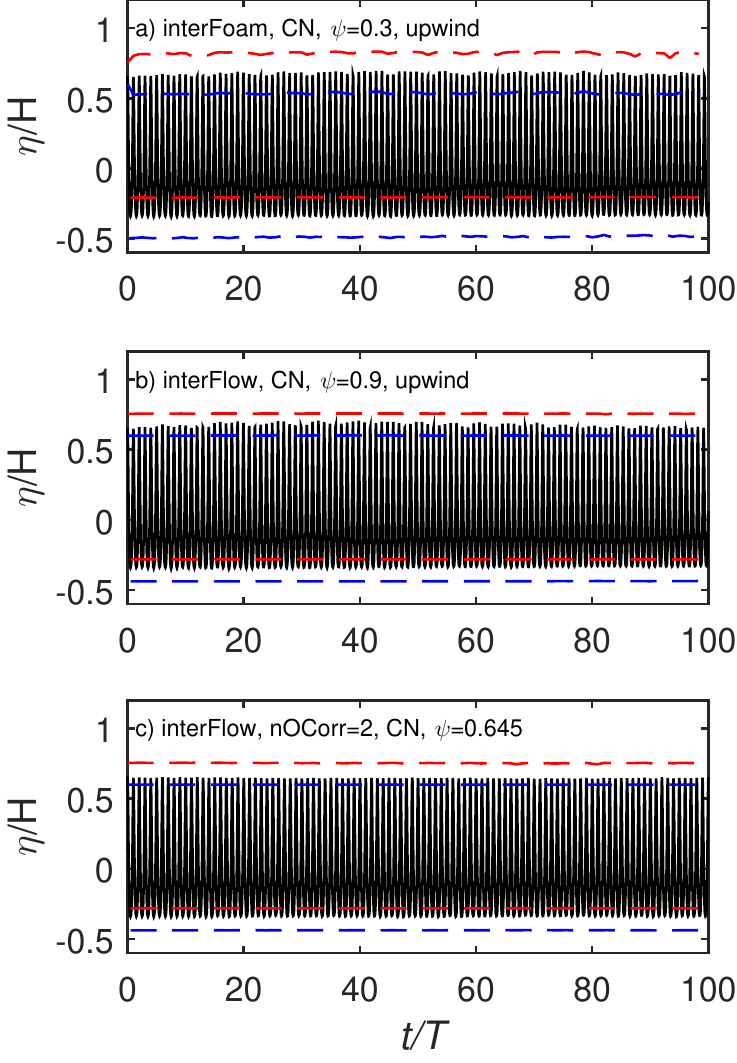}
    \cprotect\caption{Simulated surface elevations (-) as a function of time utilizing interFoam and interFlow together with the $\alpha=0.99$ and $\alpha=0.01$ (- -) contours (Main fixed parameters: $Co=0.15$, $N=12.5$, \verb|grad-Gauss Linear, laplacian-Gauss linear corrected|).}
    \label{fig:FlowFoametadif}
 \end{figure}

As shown with \verb|interFoam|, \verb|interFlow| is also sensitive to the setup, and the same diffusive balance that could be achieved with \verb|interFoam| can also be achieved with \verb|interFlow|. In Figure \ref{fig:FlowFoametadif} the simulated surface elevations utilizing \verb|interFoam| and \verb|interFlow| respectively are once again compared, this time utilizing schemes to achieve a diffusive balance.  It can be seen that \verb|interFlow|, like \verb|interFoam|, is capable of propagating the stream function wave for 100 periods, and that \verb|interFlow| throughout the simulation keeps a sharper interface as the $\alpha=0.01$ and $\alpha=0.99$ contours are much closer. It can also be seen that \verb|interFlow| does not have the same erratic surface elevation when utilizing two outer correctors together with a blended CN scheme, which can be explained by \verb|interFlow| not having an artificial compression term, and therefore the CN scheme does not excite any erratic behaviour  near the free surface. However like \verb|interFoam|, \verb|interFlow| is also very sensitive to the exact value of the blended CN scheme, and lowering the blending factor, i.e. going more towards the \verb|Euler| scheme made the wave heights decay, and raising it towards more pure CN made the wave heights increase.

\begin{figure}[ht]
	\centering
    \includegraphics[trim=0cm 0cm 0cm 0.5cm, clip=true, scale=1]{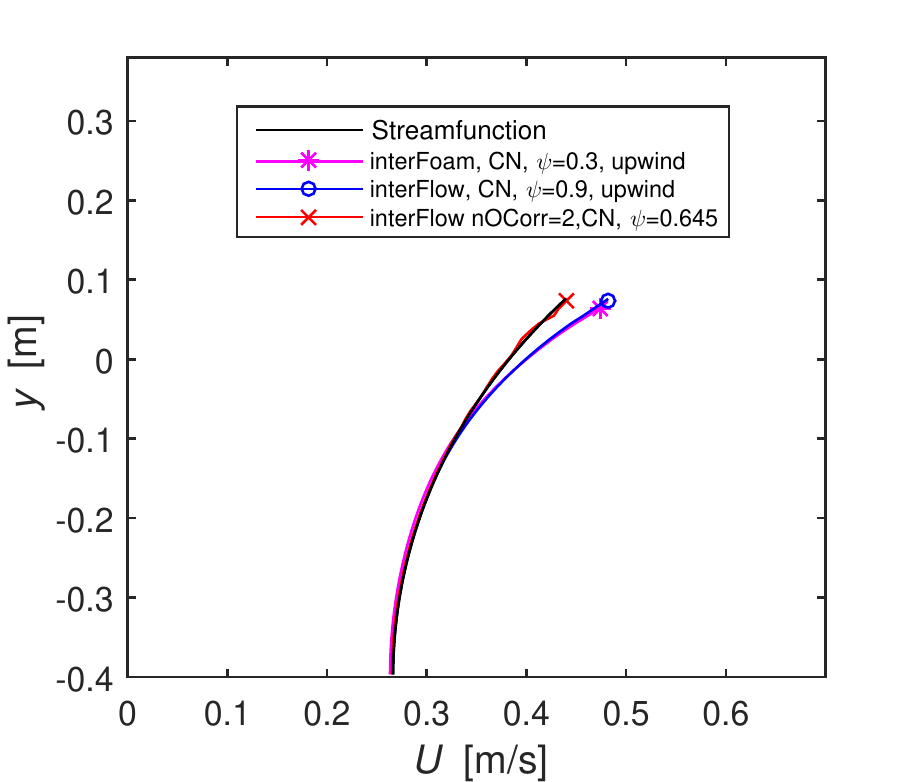}
    \cprotect\caption{Velocity distribution beneath the crest at $t=5T$ utilizing an upwind scheme for the convection for interFoam with CN($\psi$=0.3) as well as interFlow with CN($\psi$=0.9) and two outer correctors with CN($\psi$=0.645) (Main fixed parameters: $Co=0.15$, $N=12.5$, \verb|grad-Gauss Linear, laplacian-Gauss linear corrected|).}
    \label{fig:FlowFoamVelocitydiff}
 \end{figure}

The resulting velocity profiles are shown in Figure \ref{fig:FlowFoamVelocitydiff}. Here it can be seen that the two solvers perform quite similarly when utilizing an \verb|upwind| scheme together with a blended CN scheme, and that the overestimation of the velocity near the crest is reduced. Furthermore, it can be seen that \verb|interFlow| also shows a significantly improved velocity profile when switching to two outer correctors, together with a blended CN scheme and that \verb|interFlow| does not suffer, to the same degree, from oscillations in the velocity profile as did \verb|interFoam|.
 
\begin{figure}[h]
	\centering
    \includegraphics[scale=1]{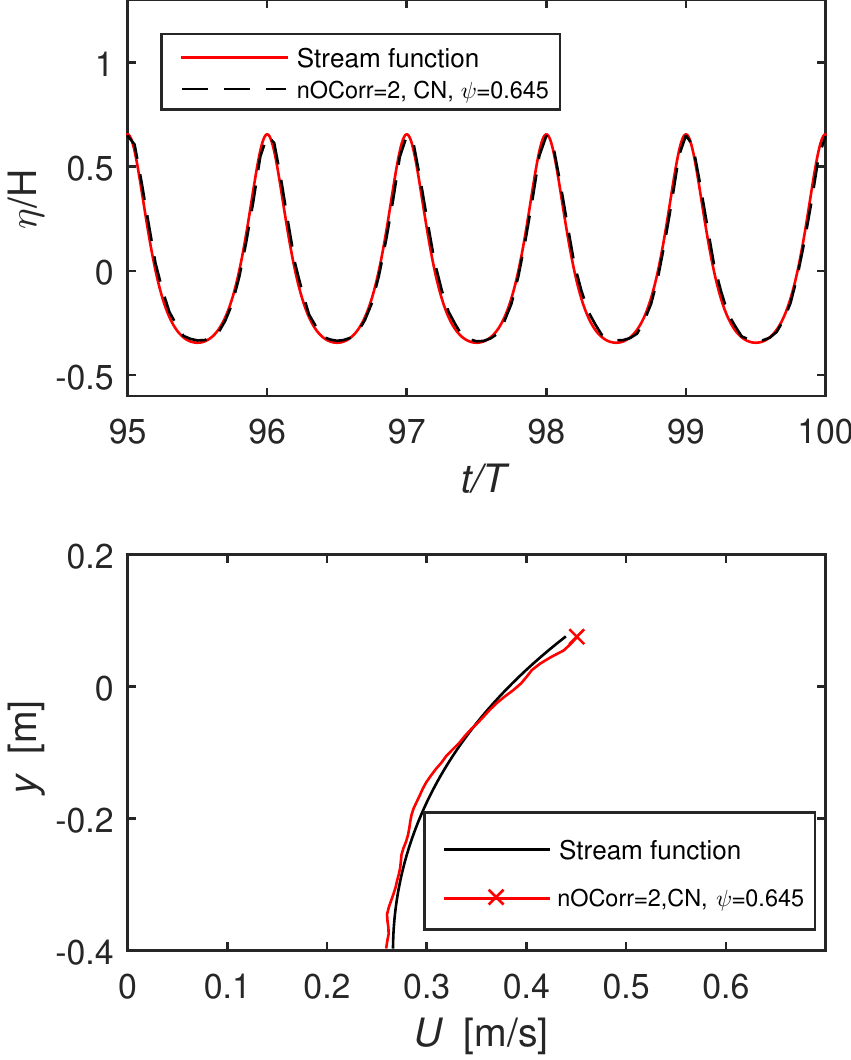}
    \caption{Surface elevations and velocity distribution beneath the crest after 100 periods utilizing interFlow.}
    \label{fig:interFlowStrength}
 \end{figure}

To further underline the impressive performance of \verb|interFlow| when utilizing a balanced setup, Figure \ref{fig:interFlowStrength} shows the surface elevation from the 95th to the 100th period together with the velocity profile beneath the crest at $t=100T$. Here it can be seen that even after propagating the nonlinear wave for 100 periods \verb|interFlow| still follows the analytical stream function solution. The surface elevations are of the right magnitude, and there are no significant phase differences. Furthermore, it can be seen that the velocity profile is likewise quite close to the analytical result, though it suffers from minor oscillations. 

\section{Conclusions}
In this study the performance of \verb|interFoam| (a widely used solver in \verb|OpenFOAM| in the simulation of progressive regular gravity waves (having intermediate depth and moderate nonlinearity) has been systematically documented. It has been shown that utilizing the basic settings of the popular \verb|interFoam| tutorial \verb|damBreak| will yield quite poor results, resulting in increasing wave heights, a wiggled interface, spurious air velocities, and severely overestimated velocities near the crest. These four problems can be reduced substantially by lowering the time step and increasing the spatial resolution. It has been shown that a rather small time step, corresponding to a Courant number $Co \approx 0.05$ is needed to give a good solution when propagating a wave even short distances of around five wave wave lengths.

To test whether an improved solution could be achieved without (drastically) lowering the time step and increasing the spatial resolution, a set of simulation have been performed, where the discretization schemes and iterative solution procedures where changed one at a time. By gradually increasing and lowering the artificial compression term ($c_\alpha$), it was identified as root of the interface wiggles, which was exacerbated when increasing the $c_\alpha$ and damped or completely removed when lowering $c_\alpha$. It was also shown how changing from first-order forward Euler time discretization scheme to the (almost) second order, and less diffusive, blended Crank-Nicolson scheme caused the wiggles to appear earlier and cover a larger part of the interface. 
The convection schemes was shown to affect not only the interface wiggles, but also the development of the wave heights as well as the velocities beneath the crest. More diffusive convection schemes removed the interface wiggles and delayed the increase in wave heights or in fact, when using an \verb|upwind| scheme, caused the wave heights to decrease. Furthermore, the more diffusive schemes also reduced the overestimation of the crest velocities. In general the effect of the gradient schemes was not as large as the convection schemes, but the \verb|fourth| scheme improved the solution, and the \verb|faceMDLimited| scheme behaved very similar to the \verb|upwind| convection scheme. Finally changing the \verb|snGrad| scheme in the Laplacian created large spurious velocities in the air phase directly above the wave. These high velocities however did not seem to influence the wave kinematics. This was further backed by simulations done with a fixed time step, which clearly indicated that the spurious air velocities, while being an unwanted and un-physical phenomenon, do not have a large impact on the wave kinematics. By combining more or less diffusive schemes it was shown that a "diffusive balance" could be reached, where it was possible to propagate the wave a full 100 wave lengths while maintaining its shape. One of these balanced settings also showed a significant improvement in the velocity profile beneath the crest.

The new open source solver \verb|interFlow| was subsequently applied, and it was shown that \verb|interFlow| was capable of propagating the wave for 100 periods. The wave decreased slightly in time, but the interface was a lot sharper, and the wiggles in surface disappeared. Regarding the velocity profile \verb|interFlow| performed slightly worse than \verb|interFoam| with the base settings. Finally it was shown that \verb|interFlow| could achieve the same kind of diffusive balance which enabled the solver to propagate the wave for 100 periods while maintaining it shape and also maintaining a good match with the analytical velocity profile.

Given its rapidly growing popularity among scientists and engineers, it is hoped that the present systematic study will raise awareness and enable users to more properly simulate a wide variety of problems involving the general propagation of surface waves within the open-source CFD package \verb|OpenFOAM|.  While the present study has focused on the canonical situation involving progressive non-breaking waves, the experience presented herein is expected to be widely relevant to other, more general, problems e.g.~involving wave-structure interactions, propagation to breaking and resulting surf zone dynamics, as well as boundary layer and sediment transport processes that result beneath surface waves, all of which fundamentally rely on an accurate description of surface waves and their underlying velocity kinematics.
\section{Acknowledgements}

The first two authors acknowledge support from the European Union project
ASTARTE--Assessment, Strategy And Risk Reduction for Tsunamis in
Europe, Grant no.~603839 (FP7-ENV-2013.6.4-3).  The third author additionally aknowledges: Sapere Aude: DFF$-$Research Talent grant from The Danish Council for
Independent Research | Technology and Production Sciences  (Grant-ID: DFF - 1337-00118). The third author also enjoys partial funding through the GTS
grant to DHI from the Danish Agency for Science, Technology and Innovation. We would like to express our sincere
gratitude for this support.

\bibliography{References} 

\begin{thebibliography}{35}
\expandafter\ifx\csname natexlab\endcsname\relax\def\natexlab#1{#1}\fi
\expandafter\ifx\csname url\endcsname\relax
  \def\url#1{\texttt{#1}}\fi
\expandafter\ifx\csname urlprefix\endcsname\relax\def\urlprefix{URL }\fi

\bibitem[{Afshar(2010)}]{Afshar2010}
Afshar, M.~A., 2010. {Numerical Wave Generation in OpenFOAM. MSc Thesis,
  Chalmers University of Technology Gotheburg, Sweden}.

\bibitem[{Brown et~al.(2016)Brown, Greaves, Magar, and Conley}]{Brown2016}
Brown, S.~A., Greaves, D.~M., Magar, V., Conley, D.~C., 2016. Evaluation of
  turbulence closure models under spilling and plunging breakers in the surf
  zone. Coast. Eng. 114, 177--193.

\bibitem[{Chen et~al.(2014)Chen, Zang, Hillis, Morgan, and
  Plummer}]{Chenetal2014}
Chen, L.~F., Zang, J., Hillis, A.~J., Morgan, G. C.~J., Plummer, A.~R., 2014.
  Numerical investigation of wave-structure interaction using openfoam. Ocean
  Eng. 88, 91--109.

\bibitem[{Deshpande et~al.(2012)Deshpande, Anumolu, and
  Trujillo}]{Deshpandeetal2012}
Deshpande, S.~S., Anumolu, L., Trujillo, M.~F., 2012. Evaluating the
  performance of the two-phase flow solver interfoam. Comput. Sci. Discov.
  5~(1), article no. 014016.

\bibitem[{Francois et~al.(2006)Francois, Cummins, Dendy, Kothe, Sicilian, and
  Williams}]{Francoisetal2006}
Francois, M.~M., Cummins, S.~J., Dendy, E.~D., Kothe, D.~B., Sicilian, J.~M.,
  Williams, M.~W., 2006. A balanced-force algorithm for continuous and sharp
  interfacial surface tension models within a volume tracking framework. J.
  Comp. Phys. 213~(1), 141--173.

\bibitem[{Galusinski and Vigneaux(2008)}]{GalusinskiVigneaux2008}
Galusinski, C., Vigneaux, P., 2008. On stability condition for bifluid flows
  with surface tension: Application to microfluidics. J. Comp. Phys. 227~(12),
  6140--6164.

\bibitem[{Greenshields(2015)}]{Greenshields2015}
Greenshields, C.~J., 2015. OpenFOAM, The Open Source CFD Toolbox, Programmer's
  Guide. Version 3.0.1. OpenFOAM Foundation Ltd.

\bibitem[{Greenshields(2016)}]{Greenshields2016}
Greenshields, C.~J., 2016. OpenFOAM, The Open Source CFD Toolbox, User's Guide.
  Version 4.0. OpenFOAM Foundation Ltd.

\bibitem[{Higuera et~al.(2013)Higuera, Lara, and Losada}]{Higueraetal2013}
Higuera, P., Lara, J.~L., Losada, I.~J., 2013. Simulating coastal engineering
  processes with {OpenFOAM (R)}. Coast. Eng. 71, 119--134.

\bibitem[{Hu et~al.(2016)Hu, Greaves, and Raby}]{Huetal2016}
Hu, Z.~Z., Greaves, D., Raby, A., 2016. Numerical wave tank study of extreme
  waves and wave-structure interaction using {OpenFoam (R)}. Ocean Eng. 126,
  329--342.

\bibitem[{Hysing(2006)}]{Hysing2006}
Hysing, S., 2006. A new implicit surface tension implementation for interfacial
  flows. Int.l J. Numer. Meth. Fluids 51~(6), 659--672.

\bibitem[{Jacobsen et~al.(2014)Jacobsen, Freds{\o}e, and
  Jensen}]{Jacobsenetal2014}
Jacobsen, N.~G., Freds{\o}e, J., Jensen, J.~H., 2014. Formation and development
  of a breaker bar under regular waves. {Part} 1: {Model} description and
  hydrodynamics. Coast. Eng. 88, 182--193.

\bibitem[{Jacobsen et~al.(2012)Jacobsen, Fuhrman, and
  Freds{\o}e}]{Jacobsenetal2012}
Jacobsen, N.~G., Fuhrman, D.~R., Freds{\o}e, J., 2012. A wave generation
  toolbox for the open-source {CFD} library: {OpenFoam (R)}. Int. J. Numer.
  Meth. Fluids 70, 1073--1088.

\bibitem[{Jacobsen et~al.(2015)Jacobsen, van Gent, and
  Wolters}]{Jacobsenetal2015}
Jacobsen, N.~G., van Gent, M. R.~A., Wolters, G., 2015. Numerical analysis of
  the interaction of irregular waves with two dimensional permeable coastal
  structures. Coast. Eng. 102, 13--29.

\bibitem[{Liu and Hinrichsen(2014)}]{LiuHinrichsen2014}
Liu, Y., Hinrichsen, O., 2014. Cfd modeling of bubbling fluidized beds using
  openfoam (r) : Model validation and comparison of tvd differencing schemes.
  Computers and Chem. Eng. 69, 75--88.

\bibitem[{Lupieri and Contento(2015)}]{LupieriContero2015}
Lupieri, G., Contento, G., 2015. Numerical simulations of 2-d steady and
  unsteady breaking waves. Ocean Eng. 106, 298--316.

\bibitem[{Meier et~al.(2002)Meier, Yadigaroglu, and Smith}]{Meieretal2002}
Meier, M., Yadigaroglu, G., Smith, B.~L., 2002. A novel technique for including
  surface tension in plic-vof methods. European J. Mech. B-fluids 21~(1),
  61--73.

\bibitem[{Menard et~al.(2007)Menard, Tanguy, and Berlemont}]{Menardetal2007}
Menard, T., Tanguy, S., Berlemont, A., 2007. Coupling level set/vof/ghost fluid
  methods: Validation and application to 3d simulation of the primary break-up
  of a liquid jet. Int. J. Multiphase Flow 33~(5), 510--524.

\bibitem[{Paulsen et~al.(2014)Paulsen, Bredmose, Bingham, and
  Jacobsen}]{Paulsenetal2014}
Paulsen, B.~T., Bredmose, H., Bingham, H.~B., Jacobsen, N.~G., 2014. Forcing of
  a bottom-mounted circular cylinder by steep regular water waves at finite
  depth. J. Fluid Mech. 755, 1--34.

\bibitem[{Popinet and Zaleski(1999)}]{PopinetZaleski1999}
Popinet, S., Zaleski, S., 1999. A front-tracking algorithm for accurate
  representation of surface tension. Int. J. Numer. Meth. in Fluids 30~(6),
  775--793.

\bibitem[{Rienecker and Fenton(1981)}]{RieneckerFenton1981}
Rienecker, M., Fenton, J., 1981. A fourier approximation method for steady
  water-waves. J. Fluid Mech. 104~(MAR), 119--137.

\bibitem[{Roenby et~al.(2016)Roenby, Bredmose, and Jasak}]{Roenbyetal2016}
Roenby, J., Bredmose, H., Jasak, H., 2016. A computational method for sharp
  interface advection. Royal Society Open Science 3~(11), 160405.

\bibitem[{Roenby et~al.(2017{\natexlab{a}})Roenby, Bredmose, and
  Jasak}]{Roenbyetal2017}
Roenby, J., Bredmose, H., Jasak, H., 2017{\natexlab{a}}. Isoadvector: Geometric
  {VOF} on general meshes. In: 11th OpenFOAM Workshop. 11th OpenFOAM Workshop.
  Springer Nature. Status: Accepted.

\bibitem[{Roenby et~al.(2017{\natexlab{b}})Roenby, Larsen, Bredmose, and
  Jasak}]{Roenbyetal2017a}
Roenby, J., Larsen, B., Bredmose, H., Jasak, H., 2017{\natexlab{b}}. A new
  volume-of-fluid method in openfoam. In: 7th Int. Conf. Comput. Methods Marine
  Eng. 7th Int. Conf. Comput. Methods Marine Eng. Nantes, France, pp. 1--12.

\bibitem[{Rudman(1997)}]{Rudman1997}
Rudman, M., 1997. Volume-tracking methods for interfacial flow calculations.
  Int.l J. Numer. Meth. Fluids 24~(7), 671--691.

\bibitem[{Schmitt and Elsaesser(2015)}]{SchmittElsaesser2015}
Schmitt, P., Elsaesser, B., 2015. On the use of {OpenFOAM} to model oscillating
  wave surge converters. Ocean Eng. 108, 98--104.

\bibitem[{Shirani et~al.(2005)Shirani, Ashgriz, and
  Mostaghimi}]{Shiranietal2005}
Shirani, E., Ashgriz, N., Mostaghimi, J., 2005. Interface pressure calculation
  based on conservation of momentum for front capturing methods. J. Comp. Phys.
  203~(1), 154--175.

\bibitem[{Tanguy et~al.(2007)Tanguy, Menard, and Berlemont}]{Tanguyetal2007}
Tanguy, S., Menard, T., Berlemont, A., 2007. A level set method for vaporizing
  two-phase flows. J. Comp. Phys. 221~(2), 837--853.

\bibitem[{Ting and Kirby(1994)}]{TingKirby1994}
Ting, F. C.~K., Kirby, J.~T., 1994. Observation of undertow and turbulence in a
  laboratory surf zone. Coast. Eng. 24~(1-2), 51--80.

\bibitem[{Tomaselli(2016)}]{Tomaselli2016}
Tomaselli, P., 2016. Detailed analyses of breaking wave dynamics interaction
  with nearshore and offshore structures. Ph.D. thesis, . Technical University
  of Denmark., Kgs. Lyngby, Denmark.

\bibitem[{Vukcevic(2016)}]{Vukcevic2016}
Vukcevic, V., 2016. Numerical modelling of coupled potential and viscous flow
  for marine applications. Ph.D. thesis, University of Zagreb, Zagreb, Croatia.

\bibitem[{Vukcevic et~al.(2016)Vukcevic, Jasak, and
  Malenica}]{Vukcevic2016etal}
Vukcevic, V., Jasak, H., Malenica, S., 2016. Decomposition model for naval
  hydrodynamic applications, part i: Computational method. Ocean Eng. 121,
  37--46.

\bibitem[{Weber(2016)}]{Weber2016}
Weber, J., 2016. {Numerical Investigation on the Damping of Water Waves in a
  Towing Tank. BSc. Thesis.Thechnical University of Berlin, Berlin, Germany. }.

\bibitem[{Wemmenhove et~al.(2015)Wemmenhove, Luppes, Veldman, and
  Bunnik}]{Wemmenhoveetal2015}
Wemmenhove, R., Luppes, R., Veldman, A. E.~P., Bunnik, T., 2015. Numerical
  simulation of hydrodynamic wave loading by a compressible two-phase flow
  method. Computers and Fluids 114, 218--231.

\bibitem[{Wroniszewski et~al.(2014)Wroniszewski, Verschaeve, and
  Pedersen}]{Wroniszewsketal2014}
Wroniszewski, P.~A., Verschaeve, J. C.~G., Pedersen, G.~K., 2014. Benchmarking
  of navier-stokes codes for free surface simulations by means of a solitary
  wave. Coast. Eng. 91, 1--17.

\end{thebibliography}


%

\end{document}